\documentclass[prl
    aps,
    twocolumn,
    superscriptaddress
]{revtex4-2}

\usepackage{amsmath,amssymb}

\usepackage{graphicx}% Include figure files
\usepackage{dcolumn}% Align table columns on decimal point
\usepackage{bm}% bold math
\usepackage{xcolor}
\usepackage{xr}
\PassOptionsToPackage{hyphens}{url}
\usepackage{hyperref}
\usepackage{url}
\usepackage{xurl}

\usepackage{enumitem}
\usepackage{accents}
\usepackage{comment}
\hypersetup{colorlinks,linkcolor={blue!90!black},citecolor={blue!90!black},urlcolor={blue!90!black}}
\usepackage{graphicx}% Include figure files
\usepackage{subcaption}
\usepackage{dcolumn}% Align table columns on decimal point
\usepackage{bm}% bold math
\usepackage{xcolor}
\usepackage{xr}
\usepackage{bm}
\usepackage{amsmath}

\PassOptionsToPackage{hyphens}{url}
\usepackage{hyperref}
\usepackage{url}
\usepackage{float}
\usepackage{booktabs}

\usepackage{xurl}

\usepackage{enumitem}
\usepackage{accents}
\usepackage{comment}
\DeclareMathAlphabet\mathbfcal{OMS}{cmsy}{b}{n}

\usepackage{mathtools}
\usepackage{euscript}

\begin{document}

%\title{Collective effects during antibiotic-induced growth-rate inhibition}
\title{ From Growth Inhibition to Fluctuations: Testing an Ising Model of Bacterial Responses to Antibiotics}
%\author{}

\author{Pietro Cicuta}
\email{pc245@cam.ac.uk}
\affiliation{Cavendish Laboratory, University of Cambridge, CB3 0HE Cambridge, United Kingdom}
\author{Morten Kals}
\affiliation{Cavendish Laboratory, University of Cambridge, CB3 0HE Cambridge, United Kingdom}
\author{Nir S. Gov}
\email{nir.gov@weizmann.ac.il}
\affiliation{Department of Chemical and Biological Physics, Weizmann Institute of Science, Rehovot 7610001, 
Israel}
\affiliation{Department of Physiology, Development and Neuroscience, University of Cambridge, Cambridge CB2 3DY, United Kingdom}

\date{\today}

\begin{abstract}
Complex networks of biochemical reactions have been proposed to exhibit emergent cooperativity that can be described by an effective Ising model, with a small number of parameters encoding collective interactions. Here this mapping is experimentally tested using the response of three bacterial species to a range of antibiotics acting on different cellular targets. The Ising description is compared to the conventional Hill model, examining both the mean growth-rate response and, importantly, its population growth-rate heterogeneity (PGRH). While both models describe the mean growth-rate dependence on antibiotic concentration comparably well, we find that the Ising model provides a markedly better description of PGRH, offering an additional and more stringent test of the collective description. The fitted Ising interaction strength is systematically related to the Hill exponent and increases with the functional distance of the antibiotic target from the protein-production machinery. These results provide experimental support for describing complex biochemical reaction networks through effective cooperative Ising interactions, and suggest that the collective response of bacterial growth to antibiotics reflects the organization of the underlying metabolic network.
\end{abstract}

\maketitle

%\emph{Introduction.} 
%
There is a growing body of work studying the overall effective cooperativity in networks of chemical reactions, and how the structure of the network can come into play together with the properties of the nodes \cite{ferrell1997responses,fitzgerald2009maximally,mora2011biological,agliari2013collective,agliari2016complete,merchan2016sufficiency}. This emergent cooperativity was mapped onto a mean-field Ising model~\cite{agliari2016complete}, in which a single effective coupling parameter determines the Hill coefficient, covering from non-cooperative Michaelis–Menten kinetics up to a switch-like critical point. In the present work we explore and test this mapping using the measured growth-rate response, and its variability, of different species of bacteria to a variety of antibiotics~\cite{kals2025antibiotics}.

The effect of different antibiotics, as function of concentration, on the growth rate of several bacteria was recently explored in several recent studies~\cite{bren2023tradeoffs,vaisbourd2025principles,kals2025antibiotics}. In all these studies it is found that the growth rate decreases, as function of increasing antibiotic concentration, as expected. The growth-rate is observed to undergo a sharp drop at some critical antibiotic concentration, with the sharpness of the drop usually interpreted as an indication for the cooperativity between the metabolic nodes of the bacterial biochemical network that have been perturbed. The data across many antibiotics, with different molecular mechanisms, shows shallow responses for protein-synthesis inhibitors, and increasingly steep responses for antibiotics that act on mechanisms more remote from protein synthesis, up to membrane-synthesis inhibitors, for which the response is close to switch-like. Merchan and Nemenman \cite{merchan2016sufficiency} argue that densely interacting networks with many-body interactions can often be described well by a pairwise maximum-entropy (Ising) model. In a `systems' view of the bacterial cell the observed trend of increasing steepness of the antibiotic response curve with increasing functional distance from the protein production hub could then be interpreted as resulting from the drug perturbation cascading though increasing number of intervening levels, analogous to the signaling cascades described by Ferrell~\cite{ferrell1997responses}.
%

%Motivated by these conceptual problems, we propose  an alternative description of the bacterial growth-rate that is based on %the Ising spin formalism~\cite{pinkoviezky2018collective,khanna2025photometric}. In this model the individual ``spins" %represent functional elements in the cell, which can be either functional ($1$) or dysfunctional ($0$) due to antibiotic %binding.   The model includes explicitly the effective interactions between the functional elements, which can drive the %growth rate to undergo a discontinuous sharp decrease. 

The step-like behavior of the growth-rate as function of the antibiotic concentration is often phenomenologically fitted to a Hill function,
with differing values of the Hill exponent~\cite{kals2025antibiotics}. The Hill function is usually interpreted as an expression for the average occupation of some binding site, where the exponent $n_H$ represents the cooperativity of binding. Hence, for this situation it could be interpreted as the antibiotic binding to its target, which then diminishes the growth rate. 
While the Hill function description is simple and can be fitted to the data \cite{kals2025antibiotics}, it does not easily correspond to a mechanistic description of the process \cite{Garcia2025}. 
%
%A Hill function with a Hill coefficient greater than $1$ is furthermore conceptually problematic when interpreted as a %literal equilibrium production law because such coefficients imply cooperative or highly nonlinear behavior that usually %cannot arise from simple independent equilibrium binding alone \cite{Bottani2017}. Only monomeric or dimeric binding (leading %to $n=1$ or $n=2$) are common molecular mechanism in cells and well characterized theoretically
%\cite{bintu2005transcriptional}. 
Large Hill coefficients produce ultrasensitive, switch-like responses that often require hidden mechanisms such as allosteric coupling, multistep regulation, feedback, or energy-consuming nonequilibrium processes~\cite{goldbeter1981amplified,altszyler2017}, rather than passive thermodynamic equilibrium. In addition, the Hill coefficient is generally only an effective steepness parameter that does not directly correspond to a specific number of binding sites or molecules.

\begin{figure*}[t!]
    %\centering
    \includegraphics[
        width=\textwidth,
        height=0.62\textheight,
        keepaspectratio
    ]{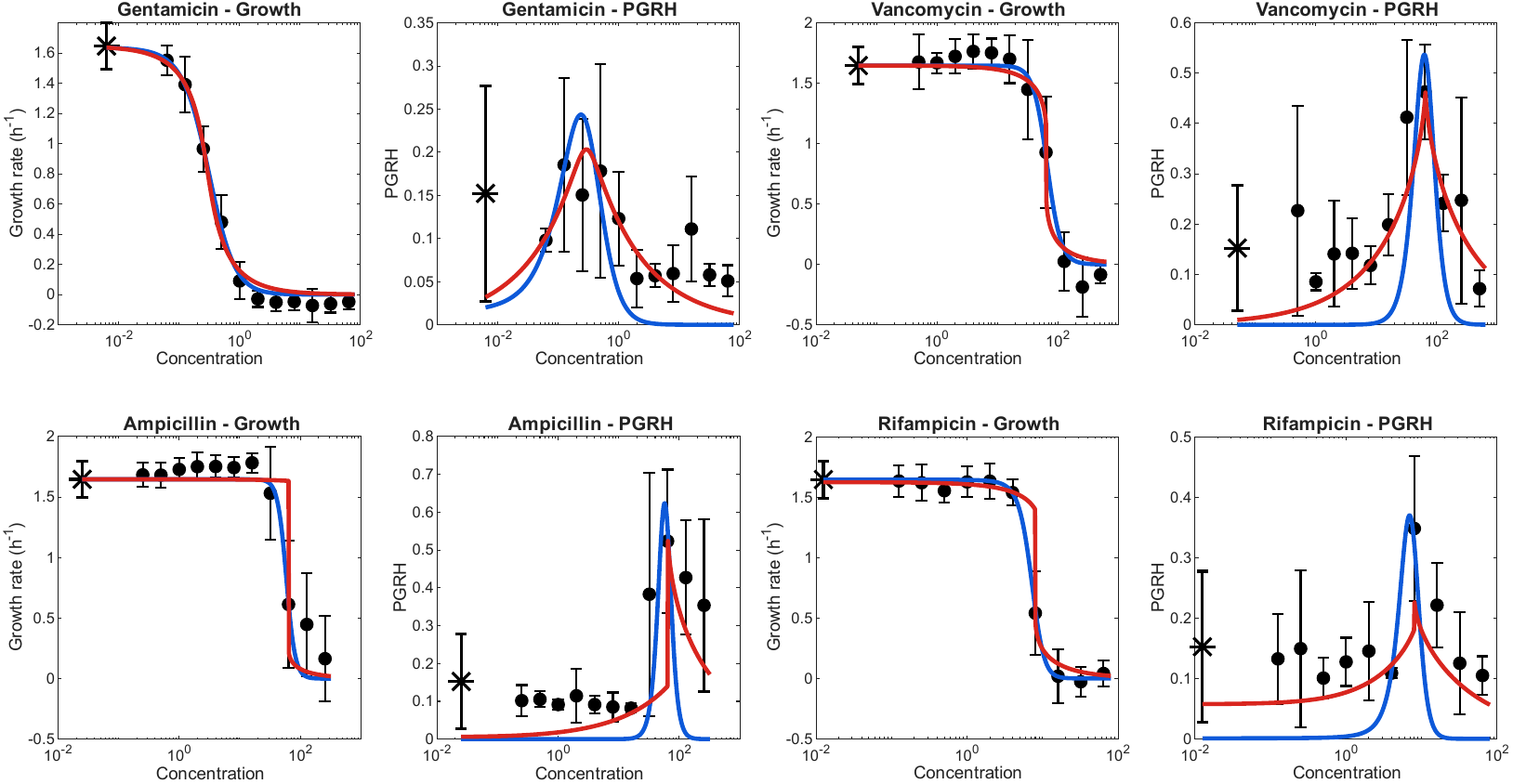}
    \caption{Comparison of Hill (blue) and Ising model (red) fits for Gentamicin, Vancomycin, Ampicillin, and Rifampicin, for \emph{E. coli}.
    For each antibiotic, the left panel shows the mean growth rate and the right panel shows PGRH.
    Blue curves represent the Hill model, while red curves represent the Ising model. Experimental measurements are shown in black. The star symbol denote the value at zero antibiotic concentration, plotted at a tenth of the minimal concentration data point in each case. The fits to the other antibiotics are shown in Figs.~S4, S5.   \label{fig:HillIsingComparison}} 
\end{figure*}

In addition to the growth rate, the population
growth rate heterogeneity (PGRH) was measured in~\cite{kals2025antibiotics}, where PGRH is defined as the standard deviation of  growth rates across many small clonal colonies. %at a given time point, averaged across multiple time points. 
There is currently no established theoretical description of the growth rate heterogeneity (PGRH), as function of the antibiotic concentration.

In this work we consider this large dataset of bacterial responses, in terms of both Ising cooperativity or Hill descriptions. We show that (1) both descriptions capture the inflection region of the growth-rate very well; (2) the descriptions are distinguishable, and the analytical link between the Hill and Ising parameters holds as a lower bound in the experimental data; (3) the variability in the growth-rate arises naturally in the Ising framework, and the PGRH data is described better by the Ising model than it is by assuming concentration fluctuations in the background of a Hill description.

%Finally, using an equilibrium occupancy formula to describe a production rate assumes strong timescale separation and a %direct mapping from binding probability to biochemical flux, even though production itself is an intrinsically dynamical and %irreversible process. Hence the $n\gg2$ observed in \cite{kals2025antibiotics} implies this Hill function is being used  %empirically, and not mechanistically, to fit the data. 

%Motivated by these conceptual problems, we propose below another description of the bacterial growth-rate that is based on the Ising spin %formalism \cite{pinkoviezky2018collective,khanna2025photometric}. In this model the individual "spins" represent functional elements in the cell, %which can be either functional ($1$) or dysfunctional ($0$), due to antibiotic binding. 
%

%

%{
%%We show in this work that using this effective Ising model allows to describe the experimentally observed growth-rates, and %provides a parameter-free prediction of the PGRH, as a  function of the antibiotic concentration. 
%}

\emph{Growth-rate heterogeneity due to concentration fluctuations.} One natural source of heterogeneity can arise from fluctuations in the concentration of antibiotics inside the cells. The concentration fluctuations can arise from fluctuations in the number of channels through which the antibiotics enter the cell, the activity of pumps that transports the antibiotics out of the cell, as well as processes that sequester the antibiotic molecules (removing them from the cytoplasm).

Antibiotic concentration fluctuations result in the following estimate for the variance in the growth rate
\begin{equation} \label{hillnoise_linear}
    \delta g(c)=|g(c)'|\delta c
\end{equation}
where $g(c)$ is the average growth-rate at the antibiotic concentration $c$,  $g(c)'=dg(c)/dc$ and $\delta c$ is the variance in the concentration. Concentration fluctuations usually follow Poissonian statistics, such that $\delta c\propto \sqrt{c}$. We therefore expect the PGRH that is driven by concentration fluctuations to behave as
\begin{equation} \label{hillnoise}
    \delta g(c)=A|g(c)'|\sqrt{c}
\end{equation}
Where $A$ is an arbitrary multiplication factor.

The growth-rate is described here by a modified Hill function, where we included an effective ``background concentration", such that: $c\rightarrow c+c_0$ 
\begin{equation} \label{hill1}
    g(c)=G(0)\frac{1}{((c+c_0)/c_{s})^n+1}
\end{equation}
This shift describes a small background concentration of inhibitory effectors that act to limit the growth-rate even in the absence of added antibiotics. This allows the calculated PGRH to have a slower decay in the limit of zero antibiotic concentration, and gives a better fit to the experimental data.

\emph{Ising model for the growth-rate and PGRH.} Motivated by the shortcomings of the phenomenological description of the growth-rate dependence on the antibiotic concentration in terms of Hill function, we explore an alternative description of the growth rate and its variance. We consider the biochemical network to be described by a finite number $N$ of interacting nodes. The complex cascade of interaction is mapped to an effective interaction between nodes, described by a pair-wise Ising form~\cite{pinkoviezky2018collective,khanna2025photometric}

%The linear kinetics of the binding process of the antibiotic to the cellular %agent is mapped to a term in the Hamiltonian that depends logarithmically on %the antibiotic concentration (derivation in SI section S1). The energy %functional of the Ising model, including the interactions between the agents, %has the form

\begin{equation}
    E=-\frac{   \tilde{J} }   {N}  \sum_{i,j}n_in_j
    -H\sum_i n_i,
    \label{energy}
\end{equation}
%in which population of intracellular agents inside the cell,
where each of the nodes represents a specific molecular module, such as ion channels or ribosomes, and is modelled to switch between two activity states,  $n_i$ (for the $i$-th agent),  $n_i=0,1$ denoting the inactive/active state respectively. The first term in Eq.\ref{energy} describes the effective interactions as an all-to-all spin-spin coupling $\tilde{J}$. From now we define a reduced interaction parameter to be $J=\tilde{J}/N$.  This interaction between the active units can be mediated by complex biochemical and biomechanical circuits within the cell, through competition over limited resources or by direct protein-protein interactions (as in clusters of ion channels on the membrane). For $J>0$, there is positive feedback between the agents so that as the order parameter $m$ (the probability of an agent to be active) increases, the transition rate to state $1$ increases (Eq.S5). For $J<0$ the agents exert a negative feedback on each other.

The second term in Eq.\ref{energy} denotes the effect of the net metabolic activation field $H$ (all energies scaled by some effective thermal energy). This field describes the activating effects of metabolites (such as ATP) and other substrates that maintain activity and drive a positive $H$, which drives the agents to be in the ON ($1$) state. The binding of antibiotics decreases this field and acts to deactivate the agent, according to Eq.S3, so we write 
\begin{equation}
    H=H_0-\log\left(\alpha\cdot c\right)
    \label{Hactivation}
\end{equation}
where $c$ is the concentration of antibiotics, $\alpha$ is some constant that depends on the affinity of the antibiotics and the agent, as well as the antibiotic's potency. The activation field has a large value $H_0$ in the absence of antibiotics, such that it gives a large proportion of active agents in the absence of antibiotics.

\begin{figure}[t!]
\centering
    \includegraphics[width=0.98\columnwidth]{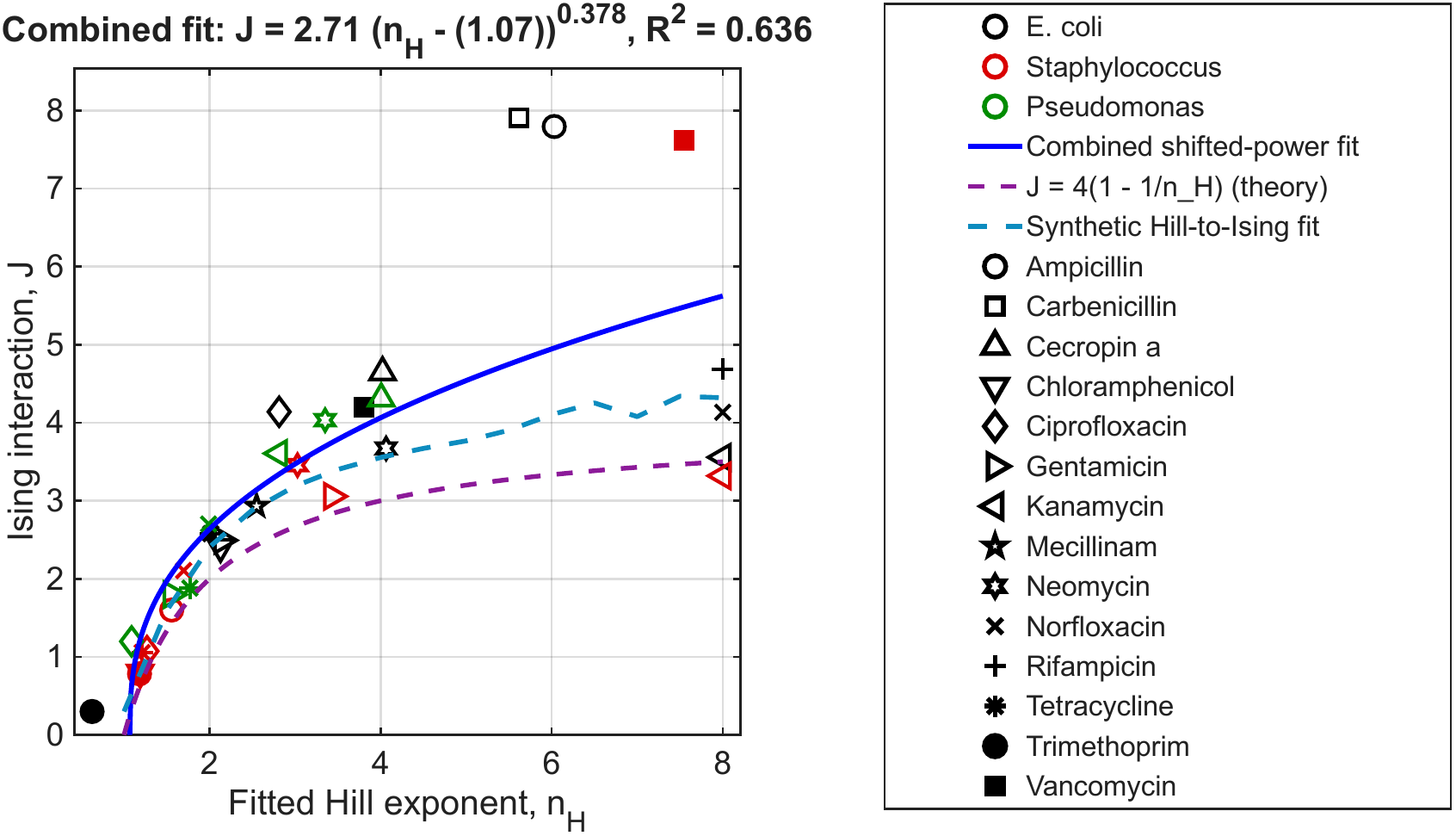}
    \caption{Comparison of fitted Hill exponents ($n_H$) and Ising interaction strengths ($J$) for \emph{E. coli} (black), \emph{Staph} (red) and \emph{Pseu} (green). The different symbols denote the different antibiotics, as given by the legend. The solid line gives a polynomial fit to the data. The purple dashed line gives the prediction of the analytic relation that matches the slope of the Hill function and Ising MF solution at the inflection point \cite{agliari2016complete} (Eq.\ref{JnH}). The blue dashed line gives the values of fitting the Ising model to exact Hill functions.}
    \label{fig:ExponentJcomb}
\end{figure}

The steady-state solution for the average probability of an agent to be active, as in the mean-field theory of ferromagnetism, is given by
\begin{equation}
    m_{MF}=\frac{1}{2}\left[1+\tanh\left(\left(Jm_{MF}+H\right)/2\right)\right].
    \label{Cself}
\end{equation}
When $J=0$, \eqref{Cself} yields the sigmoidal curve 
$m_{MF}=(1/2)(1+\tanh(H/2))$. Increasing positive $J$ produces a steeper sigmoid, until a critical value of $J_c=4$, where a discontinuity in $m_{MF}$ appears around $m_{MF}=1/2$.

We relate the the proportion of active cellular agents $m_{MF}$ to the observed growth-rate. For this comparison we have two model parameters to fit, namely $\alpha$ (Eq.\ref{Hactivation}) and $J$ (Eq.\ref{energy}). The parameter $\alpha$ simply shifts the sigmoidal solution $m_{MF}$ along the concentration axis to fit to the concentration where the growth rate has the maximal gradient, and $J$ is varied to fit to the maximal steepness of the growth rate. 

We next calculate the variance in the activity $m_{MF}$ from our spin model \cite{pinkoviezky2018collective,khanna2025photometric}
\begin{equation}
{\rm var}(m)=A\left[(1-m_{MF})w_{\text{off} \to \text{on}}+m_{MF} w_{\text{on} \to \text{off}}\right]
\label{variance}
\end{equation}
Note that this variance (Eq.\ref{variance}) is distinct from that which arises due to concentration fluctuations (Eq.\ref{hillnoise}). It arises from the dynamics of the cellular agents (spins) turning on/off which gives variations in the instantaneous number of active elements, and is therefore inversely proportional to the total number of these agents: $A\propto1/N$ \cite{pinkoviezky2018collective}. 

\emph{Comparison to experimental results.} We fitted the experimental data of the growth-rate to both the Hill model (Eq.\ref{hill1}) and the Ising model (Eq.\ref{Cself}), where we identify the average magnetization $m$ as representative of the growth rate. The fitting is done by scanning over parameter space and minimizing the global error function, as described in the SI. Note that both the Hill and Ising models have three free parameters when fitting the growth-rate data, and a single multiplication factor for the fit to the PGRH.

\begin{figure}[t!]
%\centering
    \includegraphics[width=0.98\columnwidth]{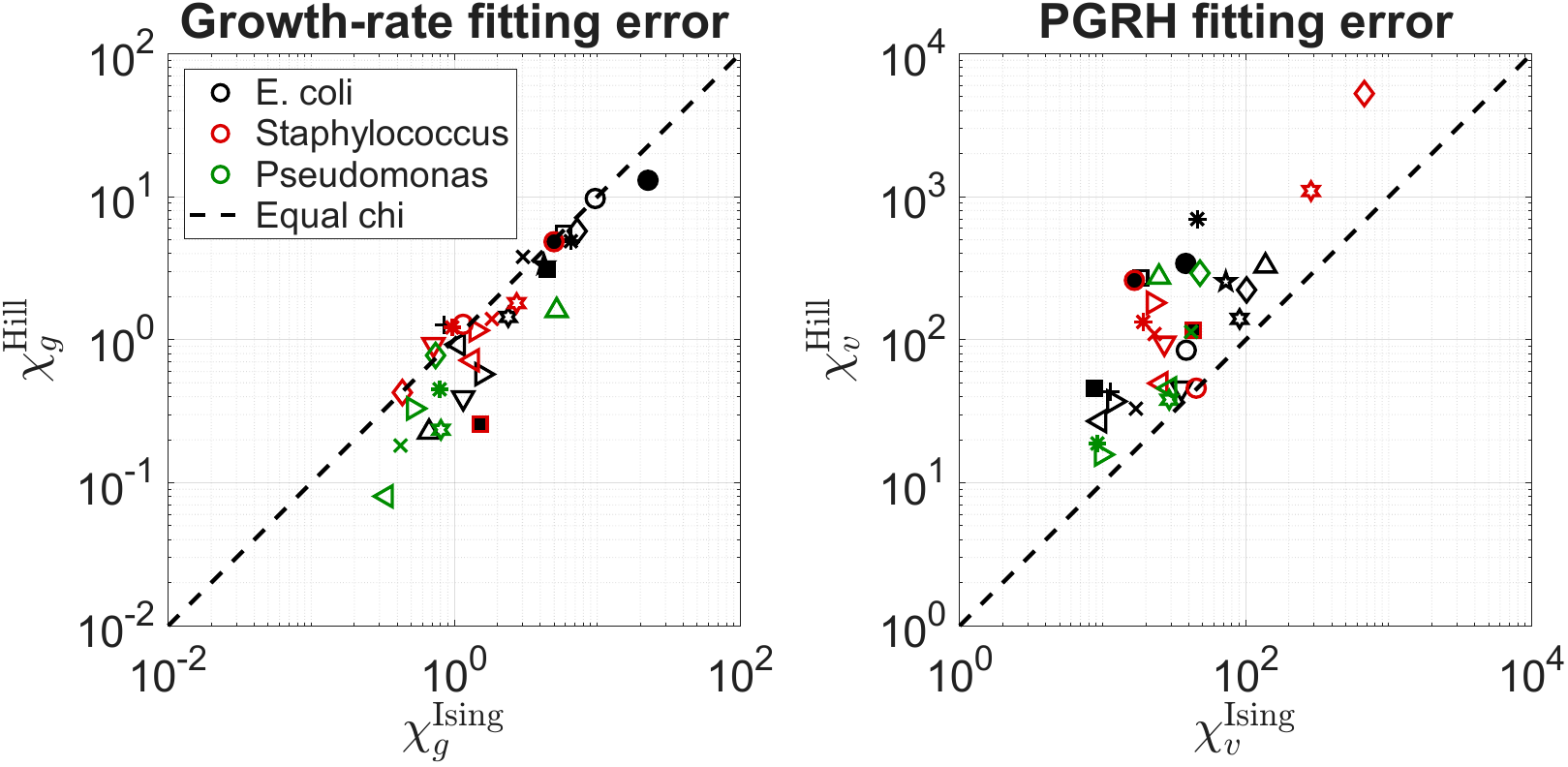}
    \caption{Comparison of fitting errors for the Hill and Ising models for the measured growth-rate (left) and the PGRH (right), for \emph{E. coli} (black), \emph{Staph} (red) and \emph{Pseu} (green). The different symbols denote the different antibiotics, as given by the legend (data symbols as in Fig.\ref{fig:ExponentJcomb}).  \label{fig:chi}}
\end{figure}

Following the fitting of the growth-rate, we fit the corresponding (species and antibiotic) PGRH data with only one free parameter, the amplitude. For the Hill function this multiplicative constant $A$ appears in Eq.\ref{hillnoise}, and for the Ising model in Eq.\ref{variance}.

In Fig.\ref{fig:HillIsingComparison} we show several typical comparisons between the Hill and Ising models for four antibiotics acting on \emph{E.coli} bacteria. These four cases span the range of fitted values of the Hill exponent and Ising interaction. The rest of the comparisons are given in Figs.S4,S5 (all the fitting parameters are given in Tables~S2,S3). 

The relationship between the fitted Hill exponent and Ising interaction strength, using all three bacteria species, is shown in Fig.~\ref{fig:ExponentJcomb} (the graphs for each species separately are given in Fig.S10). The solid line is a polynomial fit for the overall correlation, using the relation: $J=a\cdot (n_H-n_0)^b$. We find that the limit of non-interacting Ising spins ($J\rightarrow0$) corresponds to independent (linear) Hill binding kinetics ($n_H\rightarrow1$). As the Ising interaction approaches the critical value of $J\rightarrow J_c=4$, the Ising model predicts a sharp drop in the growth-rate, and indeed around this value the corresponding Hill exponents increase sharply. The correspondence between these two descriptions was previously discussed \cite{agliari2013collective}, and one can obtain a simple analytic relation if the slope of both the Hill (Eq.\ref{hill1}) and Ising (Eq.\ref{Cself}) models are equated at their inflection point, giving:
\begin{equation}
    J/J_c=1-\frac{1}{n_H}
    \label{JnH}
\end{equation}
We find that this relation behaves as a lower bound on the results we obtain from fitting the experimental data. This may indicate that to obtain a better fit to the data away from the inflection point, the Ising interaction needs to be higher than given by Eq.\ref{JnH}. This is indeed shown by fitting the Ising model to exact Hill functions (see SI section S9), which gives the points along the blue dashed line in Fig.~\ref{fig:ExponentJcomb}, which correspond very well to the experimental data fitting. A few outliers with very high fitted values of $J$ are caused by the lack of sufficiently dense experimental data near the inflection point.

Note that there are no cases where the best fit is for $J<0$, which means that all the cellular targets of the antibiotics effectively have ferromagnetic-like (mutually enhancing) interactions. Second, there is a positive correlation, whereby larger Hill exponents correspond to larger values of the Ising interaction. This means that the interpretation of large Hill exponent as an indication of cooperative interactions, agrees with the Ising model which explicitly describes them. 

Ising interactions that are larger than the critical value correspond to the abrupt drop in the growth-rate and the sharp cusp in the PGRH, as shown for example in Fig.\ref{fig:HillIsingComparison} for Ampicillin and Rifampicin. The abrupt drop occurs since the Ising model enters into a region of bistability, with two coexisting solutions. We show in the SI section S10 that the lower branch solution, which is the one shown in Fig.\ref{fig:HillIsingComparison} gives a significantly between fit to the PGRH data compared with choosing the higher branch solution (appropriately shifted along the antibiotic concentration axis to fit the sharp drop in the growth-rate, Fig.S12). We do not know if the sharp drops in the growth-rate predicted by the Ising model do actually occur in the cells, or arise here due to fitting an under-sampled experimental data set. However, there is an interesting outcome of this bistability in the Ising model, namely that the PGRH is predicted to be asymmetric around its peak, which seems to appear also in the experimental data, such as for Ampicillin (Fig.\ref{fig:HillIsingComparison}).

%\begin{figure}[t]
%\centering
%    \includegraphics[width=0.98\columnwidth]{HillExponent_IsingJ_comparison.pdf}
%    \caption{Comparison of fitted Hill exponents and Ising interaction strengths for \emph{E. coli}. The four black-filled symbols correspond to the antibiotics given in Fig.\ref{fig:HillIsingComparison}. The fits to the other antibiotics are shown in Figs.S4,S5. The dashed line is a polynomial best fit to serve as a guide of the general trend.}
%    \label{fig:ExponentJ}
%\end{figure}

In Fig.\ref{fig:chi} we plot the calculated errors for the fits using the Hill and Ising models ($\chi_g$, see SI for details), for all three bacteria and all antibiotics (the graphs for each bacteria separately are given in Fig.S13). For the growth-rate the two models give similar errors, with most of the points along the diagonal. However, for the PGRH, the Ising model gives consistently and significantly better fit of the data (Fig.\ref{fig:chi}), i.e. the shape of the `susceptibility' peak is captured better. These peaks, see Fig.\ref{fig:ExponentJcomb}, have quite different shapes and are systematically shifted in the peak position. For completeness we also tested the fit to the measured PGRH if the concentration fluctuations depend linearly on the concentration, i.e. in Eq.\ref{hillnoise} we have:$\delta g(c)=A|g(c)'|c$. This did not give significant change to the quality of the fit (Fig.S14), except for one antibiotic.

%\emph{Discussion.} 
%
%
In \cite{kals2025antibiotics} it was observed that the Hill exponent grows as the antibiotic acts on a cellular target that is further from the protein synthesis hub. We propose that this can now be interpreted with more insight into the complex networks in the cell, using the Ising model: cellular agents that are more upstream with respect to the protein production, are connected by more biochemical pathways before they affect the protein synthesis. This increase in the number of connecting biochemical pathways results in stronger effective interaction between these elements, corresponding to a larger value of the overall $J$ (Eq.\ref{energy}), which reflects a cascade of underlying transitions.

%We demonstrate that the observed growth rate heterogeneity (PGRH) could arise from both fluctuations in the concentration of %antibiotics inside the cells, as well as due to the activation dynamics of a finite population of metabolic cellular agents %(as described by the Ising model). While the Ising model description overall gives a better fit to the observed data for the %growth-rate variance, both mechanisms can contribute to the overall variations in the observed growth-rate.

A result of this study is that the effective Ising description accounts not only for the mean growth-rate response but also for the observed population growth-rate heterogeneity. While both the Hill and Ising models reproduce the average growth-inhibition curves, the measured PGRH provides a more discriminating test. In the Hill framework, heterogeneity is estimated from fluctuations in intracellular antibiotic concentration, whereas in the Ising model it emerges naturally from stochastic transitions between the activity states of interacting cellular elements. The Ising description generally provides a substantially better fit to the observed PGRH, particularly near the sharp growth-rate transition. This agreement supports the relevance of collective cellular dynamics to growth-rate fluctuations, although additional sources of heterogeneity, including intracellular antibiotic-concentration fluctuations, may also contribute.
This work 
%has shown that a large dataset of bacterial physiology can be interpreted via few parameters of an effective Ising %cooperative model, representing the structure of the bacterium's  biochemical reaction network. Future study can explore how %this effective Ising interactions between the antibiotic's metabolic cellular targets can be incorporated into a fuller model %of cell growth \cite{greulich2015growth}.
%
supports the view that collective properties of the biochemical network shape the cellular response. More broadly, it provides  a framework for connecting experimentally accessible response curves and fluctuations to the effective interactions emerging from complex intracellular networks. Future work may establish how these interactions arise from specific biochemical pathways and incorporate them into more detailed models of bacterial growth~\cite{greulich2015growth,hwa2023}.
\newpage
\textbf{Supplementary Information - From Growth Inhibition to Fluctuations: Testing an Ising Model of Bacterial Responses to Antibiotics}

\renewcommand{\thesection}{S\arabic{section}}
\renewcommand{\thefigure}{S\arabic{figure}}
\renewcommand{\thetable}{S\arabic{table}}
\renewcommand{\theequation}{S\arabic{equation}}
\setcounter{section}{0}
\setcounter{figure}{0}
\setcounter{table}{0}
\setcounter{equation}{0}
\setcounter{secnumdepth}{3}

%\author{}

\section{Mapping linear kinetics of antibiotics binding to an energy term in the Ising model}

We start with the linear kinetics of the binding process of the antibiotic to the cellular agent. Let $m(t)$ be the probability of a cellular agent to be in the ON state, equivalent to the magnetization in spin terms. Then

\[
\frac{d m}{dt}
=
- w_{\text{on} \to \text{off}} \, m
+ w_{\text{off} \to \text{on}} \, (1 - m).
\]

At steady state, ${d m}/{dt} = 0$, giving

\[
m^{\text{eq}}
=
\frac{w_{\text{off} \to \text{on}}}
{w_{\text{on} \to \text{off}} + w_{\text{off} \to \text{on}}}.
\]

The rates can be described by Glauber dynamics with attempt rate $\gamma$
\begin{equation}
w_{\text{on} \to \text{off}}
=
\frac{\gamma}{2}
\left(1 + \tanh\left(\frac{\beta \Delta E}{2}\right)\right),
\label{omega_off_bare}
\end{equation}

\begin{equation}
w_{\text{off} \to \text{on}}
=
\frac{\gamma}{2}
\left(1 - \tanh\left(\frac{\beta \Delta E}{2}\right)\right).
\label{omega_on_bare}
\end{equation}

Using an effective energy field that represents the effect of the antibiotic concentration ($c$)

\begin{equation}
\Delta E \equiv E_{\text{on}} - E_{\text{off}} = k_B T \ln\!\left(\frac{K_d}{[c]}\right).
\label{DeltaE}
\end{equation}
and
%\[
%\tanh\left(\frac{1}{2}\ln x\right) = \frac{x-1}{x+1},
%\]
%with $x = \frac{[c]}{K_d}$, x
we obtain

\[
m^{\text{eq}}
=
\frac{K_d}{K_d + [c]}
=
\frac{1}{1 + [c]/K_d}.
\]

\[
p_{\text{off}}^{\text{eq}} = 1 - m^{\text{eq}} = \frac{[c]}{K_d + [c]}.
\]

%and the rates satisfy

%\[
%\frac{w_{\text{on} \to \text{off}}}{w_{\text{off} \to \text{on}}}
%=
%\frac{[c]}{K_d}
%=
%e^{-\beta (E_{\text{off}} - E_{\text{on}})},
%\]

ensuring thermodynamic consistency.

The transition rates between the $0$ and $1$ states (using Eq.4), are now given as
\begin{equation}
w_{\text{on} \to \text{off}}
=
\frac{\gamma}{2}
\left(1 + \tanh\left(\frac{Jm+H}{2}\right)\right),
\label{omega_off}
\end{equation}

\begin{equation}
w_{\text{off} \to \text{on}}
=
\frac{\gamma}{2}
\left(1 - \tanh\left(\frac{Jm+H}{2}\right)\right).
\label{omega_on}
\end{equation}
Here, $J$ and $H$ are dimensionless variables corresponding to $\tilde{J}$ and $\tilde{H}$
scaled by some effective thermal energy.

\section{Curve fitting to the Hills model}
\subsection{Fitting procedure for the Hill growth model}

For each antibiotic, the concentration-dependent mean growth rate and growth-rate heterogeneity can be fitted simultaneously using a Hill-type response model. The model for the mean growth rate was defined as
\begin{equation}
g(c)
=
\frac{g_0}
{1+\left(\dfrac{c+c_0}{c_{\mathrm{H}}}\right)^{n_{\mathrm{H}}}},
\end{equation}
where $c$ is the antibiotic concentration, $g_0$ is the experimentally measured mean growth rate at zero antibiotic concentration, $c_{\mathrm{H}}$ is the characteristic Hill concentration, $n_{\mathrm{H}}$ is the Hill coefficient, and $c_0$ is a concentration offset.

The three nonlinear parameters fitted to the data were therefore
\begin{equation}
c_{\mathrm{H}}, \qquad n_{\mathrm{H}}, \qquad c_0.
\end{equation}

The growth-rate heterogeneity was modeled from the concentration derivative of the Hill curve. The derivative is
\begin{equation}
\frac{dg}{dc}
=
-
\frac{
g_0 n_{\mathrm{H}}
(c+c_0)^{n_{\mathrm{H}}-1}
}{
c_{\mathrm{H}}^{n_{\mathrm{H}}}
\left[
1+\left(\dfrac{c+c_0}{c_{\mathrm{H}}}\right)^{n_{\mathrm{H}}}
\right]^2
}.
\end{equation}

The predicted heterogeneity was taken to be proportional to the magnitude of this derivative multiplied by the concentration fluctuations which are proportional to $\sqrt{c+c_0}$,
\begin{equation}
v_{\mathrm{model}}(c)
=
A_{\mathrm{H}}
\left|
\frac{dg}{dc}
\right|
\sqrt{c+c_0},
\end{equation}
where $A_{\mathrm{H}}$ is an arbitrary multiplicative scale factor. For each trial set of $(c_{\mathrm{H}},n_{\mathrm{H}},c_0)$, the optimal value of $A_{\mathrm{H}}$ was calculated analytically by weighted least squares and was therefore not scanned as an additional nonlinear parameter.

\subsection{Objective function}

The mean growth-rate and heterogeneity data were fitted simultaneously by minimizing a combined weighted error function.

For the mean growth-rate data, the normalized residual at concentration $c_i$ was defined as
\begin{equation}
R_{g,i}
=
\frac{
g_i-g_{\mathrm{model},i}
}{
\sigma_{g,i}
},
\end{equation}
where $g_i$ is the experimentally measured mean growth rate and $\sigma_{g,i}$ is the experimental uncertainty assigned to the growth-rate measurement.

Each growth-rate residual was additionally weighted by the magnitude of the measured growth rate,
\begin{equation}
w_{g,i}
=
\frac{|g_i|}
{\left\langle |g| \right\rangle}.
\end{equation}

where
\begin{equation}
\langle |g| \rangle =
\frac{1}{N}\sum_{j=1}^{N}|g_j|,
\end{equation}
and $N$ is the total number of measured concentration points,
including zero concentration. This normalization ensures that
the average weight is unity, giving greater importance to
measurements with higher growth rates.

The growth-rate contribution to the objective error function was therefore
\begin{equation}
\chi_g
=
\sum_i
w_{g,i} R_{g,i}^2.
\end{equation}

For the growth-rate heterogeneity data, the normalized residual was
\begin{equation}
R_{v,i}
=
\frac{
v_i-v_{\mathrm{model},i}
}{
\sigma_{v,i}
},
\end{equation}
where $v_i$ is the measured growth-rate heterogeneity and $\sigma_{v,i}$ is its experimental uncertainty.

The he heterogeneity residuals were weighted by the measured heterogeneity data,
\begin{equation}
w_{v,i}
=
\frac{|v_i|}
{\left\langle |v| \right\rangle}.
\end{equation}

The heterogeneity contribution was then
\begin{equation}
\chi_v
=
\sum_i
w_{v,i} R_{v,i}^2.
\end{equation}

The most general total objective function minimized by the fit can be written as
\begin{equation}
\chi_{\mathrm{tot}}
=
W_g\chi_g
+
W_v\chi_v,\label{chi_hill}
\end{equation}
However, we used only the growth-rate weighting
\begin{equation}
W_g = 1,
\qquad
W_v = 0.
\end{equation}

\begin{figure}[p]
    \centering

    \includegraphics[
        width=\textwidth,
        height=0.85\textheight,
        keepaspectratio
    ]{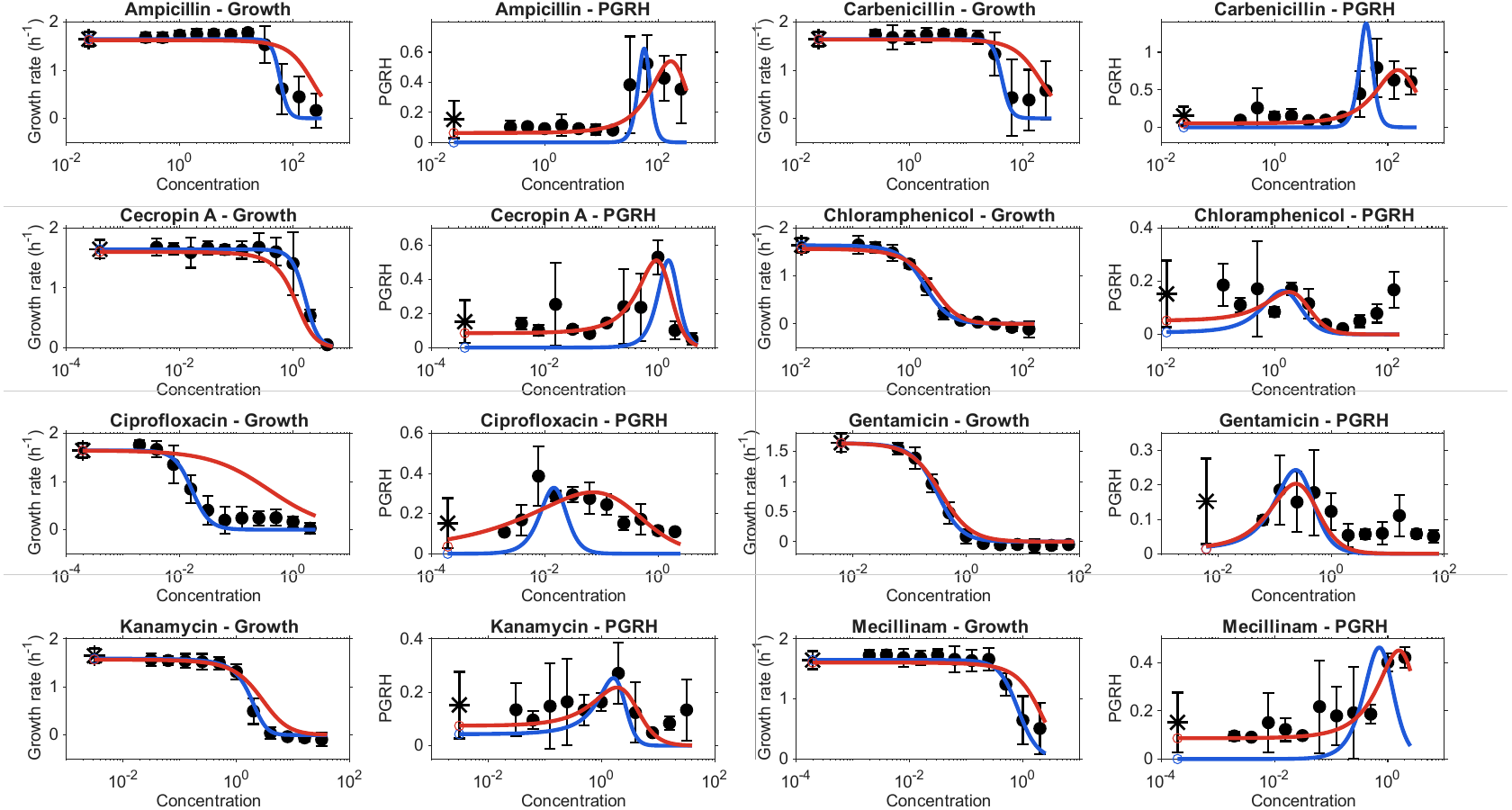}

    \caption{
    \textbf{Comparison of Hill-model fits to bacterial growth
    rates and growth-rate heterogeneity (first eight antibiotics).}
    For each antibiotic, the left panel shows the mean growth
    rate as a function of antibiotic concentration, while the
    right panel shows the population growth-rate heterogeneity
    (PGRH).
    Black symbols represent the experimental measurements,
    with error bars indicating the experimental uncertainties.
    Blue solid curves correspond to fits obtained by minimizing
    only the growth-rate error ($W_g=1$, $W_v=0$),
    whereas red solid curves correspond to simultaneous fitting
    of growth rate and PGRH ($W_g=1$, $W_v=1$).
    }

    \label{fig:HillSupplementary1}
\end{figure}

\begin{figure}[p]
    \centering

    \includegraphics[
        width=\textwidth,
        height=0.85\textheight,
        keepaspectratio
    ]{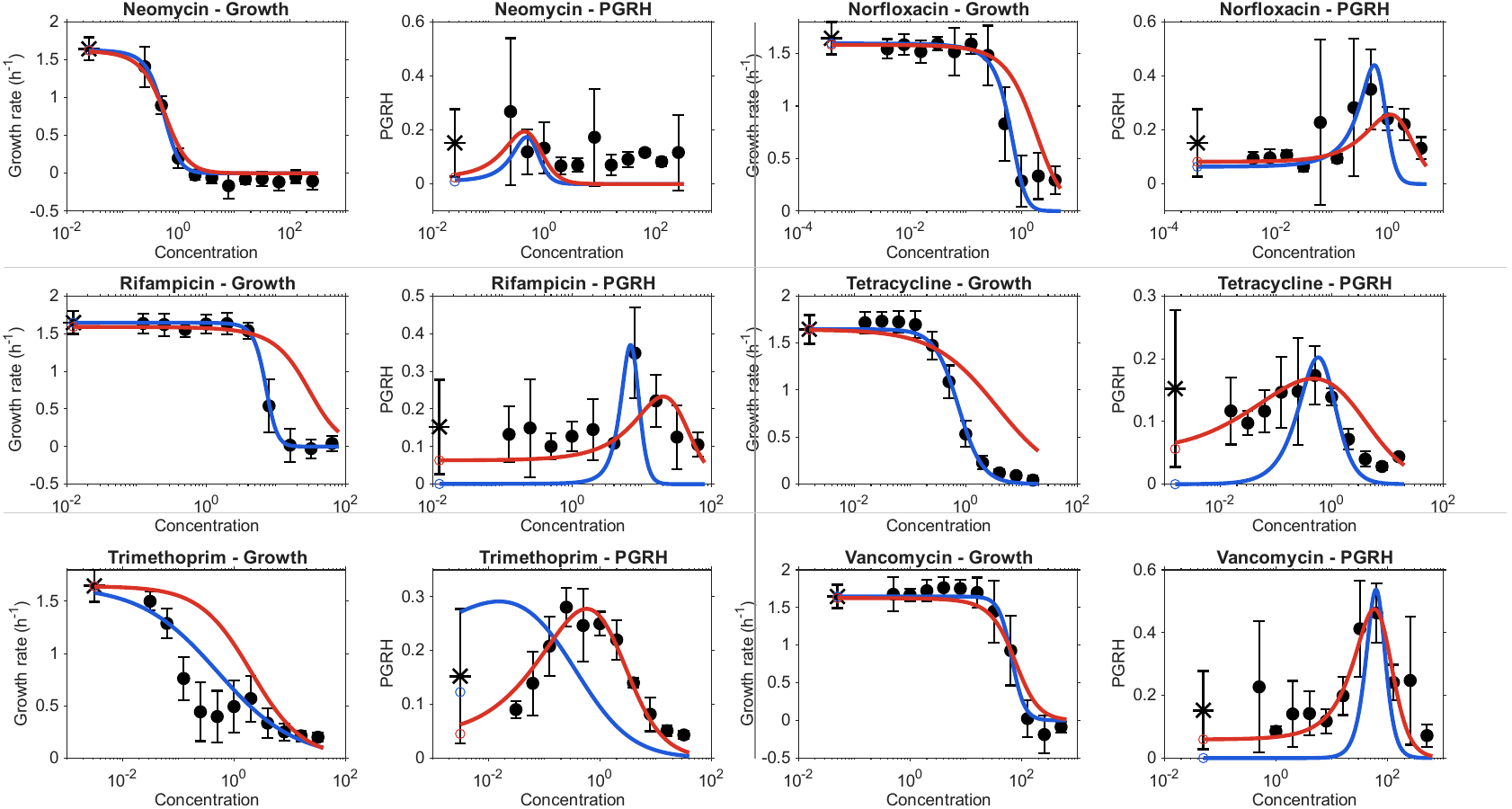}

    \caption{
    \textbf{Comparison of Hill-model fits to bacterial growth
    rates and growth-rate heterogeneity (remaining six antibiotics).}
    The organization and symbols are the same as in
    Fig.~\ref{fig:HillSupplementary1}.
    Each antibiotic is represented by two panels:
    mean growth rate (left) and PGRH (right).
    Blue curves show fits using only the mean growth-rate data
    ($W_g=1$, $W_v=0$), while red curves show joint fits
    to growth rate and heterogeneity ($W_g=1$, $W_v=1$).
    }

    \label{fig:HillSupplementary2}
\end{figure}

Other weighting, for example giving equal weight to both data sets, led to very large errors in fitting the growth-rate data, for which there is much greater confidence. This is shown in Figs.\ref{fig:HillSupplementary1},\ref{fig:HillSupplementary2}.

% ==========================================================
% Supplementary Figure S2: Remaining 6 antibiotics
% ==========================================================

\subsection{Determination of the heterogeneity scale factor}

For a given set of Hill parameters, define
\begin{equation}
s_i
=
\left|
\frac{dg}{dc}
\right|_{c_i}
\sqrt{c_i+c_0}.
\end{equation}

The heterogeneity model is then
\begin{equation}
v_{\mathrm{model},i}
=
A_{\mathrm{H}} s_i.
\end{equation}

For each trial set of $(c_{\mathrm{H}},n_{\mathrm{H}},c_0)$, the value of $A_{\mathrm{H}}$ minimizing the weighted heterogeneity error was calculated analytically as
\begin{equation}
A_{\mathrm{H}}
=
\frac{
\sum_i
\left(
\dfrac{w_{v,i}}{\sigma_{v,i}^2}
\right)
s_i v_i
}{
\sum_i
\left(
\dfrac{w_{v,i}}{\sigma_{v,i}^2}
\right)
s_i^2
}.
\end{equation}

The fitted scale factor was restricted to be non-negative.

\subsection{Initial parameter scan}

An initial three-dimensional scan was performed over $c_{\mathrm{H}}$, $n_{\mathrm{H}}$, and $c_0$.

Because the concentration ranges differed substantially between antibiotics, the scan limits for $c_{\mathrm{H}}$ and $c_0$ were defined separately for each antibiotic. Let $c_{\min}$ denote the smallest non-zero experimental concentration and $c_{\max}$ the largest experimental concentration.

The Hill concentration was scanned over
\begin{equation}
0.1\,c_{\min}
\leq
c_{\mathrm{H}}
\leq
2\,c_{\max}.
\end{equation}
A total of 55 logarithmically spaced values of $c_{\mathrm{H}}$ were evaluated.

The Hill coefficient was scanned linearly over
\begin{equation}
0.3
\leq
n_{\mathrm{H}}
\leq
8,
\end{equation}
using 55 equally spaced values. The corresponding spacing was
\begin{equation}
\Delta n_{\mathrm{H}}
=
\frac{8-0.3}{55-1}
\simeq 0.1426.
\end{equation}

The concentration offset was scanned over
\begin{equation}
10^{-4}c_{\min}
\leq
c_0
\leq
c_{\max},
\end{equation}
using 55 logarithmically spaced values.

The initial coarse scan therefore evaluated
\begin{equation}
55^3
=
166375
\end{equation}
parameter combinations for each antibiotic.

For every parameter combination, the Hill growth curve and its analytical derivative were calculated at all experimental concentrations, the optimal heterogeneity scale factor $A_{\mathrm{H}}$ was determined analytically, and the combined objective function $\chi_{\mathrm{tot}}$ was evaluated.

The parameter combination with the smallest value of $\chi_{\mathrm{tot}}$ was retained as the best result of the coarse scan.

\subsection{Continuous parameter refinement}

The best point from the coarse parameter scan was subsequently used as the starting point for a continuous local refinement using MATLAB's \texttt{fminsearch} routine.

All three positive Hill parameters were optimized in logarithmic coordinates,
\begin{equation}
p_1 = \ln c_{\mathrm{H}},
\qquad
p_2 = \ln n_{\mathrm{H}},
\qquad
p_3 = \ln c_0.
\end{equation}

This parametrization guarantees that all fitted parameters remain positive during the optimization.

The continuous refinement was restricted to remain inside the parameter ranges used in the initial scan. Trial values outside these limits were assigned a large penalty in the objective function.

At every step of the refinement, the Hill curve was recalculated, its analytical derivative was evaluated, the optimal heterogeneity scale factor $A_{\mathrm{H}}$ was recalculated analytically, and the complete weighted objective function was evaluated.

The final reported values of
\begin{equation}
c_{\mathrm{H}},
\qquad
n_{\mathrm{H}},
\qquad
c_0,
\qquad
A_{\mathrm{H}}
\end{equation}
therefore correspond to the refined minimum of the combined growth-rate and heterogeneity objective function, rather than being restricted to the discrete values used in the initial parameter scan.

\section{Curve fitting to the Ising model}

\subsection{Fitting procedure for the Ising-like growth model}

For each antibiotic, the concentration-dependent mean growth rate and growth-rate heterogeneity were fitted simultaneously using an Ising-like model. The model is controlled by three parameters, the interaction strength $J_0$, the concentration-scaling parameter $\alpha$, and the concentration offset $c_0$.

For a given antibiotic concentration $c$, the effective field was defined as
\begin{equation}
h(c) = 4 - \ln\left[(c+c_0)\alpha\right].
\end{equation}

The magnetization $m$ was obtained numerically from the steady-state condition of the Ising-like dynamics. The transition rates were
\begin{equation}
r_{+} = \frac{1}{1+\exp[-(J_0m+h)]},
\end{equation}
and
\begin{equation}
r_{-} = \frac{1}{1+\exp[J_0m+h]}.
\end{equation}

The corresponding dynamical equation is
\begin{equation}
\frac{dm}{dt}
=
(1-m)r_{+}-mr_{-}.
\end{equation}

At steady state, this reduces to
\begin{equation}
m=r_{+}.
\end{equation}

The steady-state magnetization was therefore calculated iteratively at each drug concentration. The iteration was started from $m=0$ and continued until the change in magnetization was smaller than $10^{-10}$, with a maximum of 5000 iterations. A damping factor of 0.5 was used to improve numerical stability.

The predicted mean growth rate was taken to be
\begin{equation}
g_{\mathrm{model}}(c)=g_0 m(c),
\end{equation}
where $g_0$ is the experimentally measured mean growth rate at zero antibiotic concentration.

The model prediction for the growth-rate heterogeneity was calculated from
\begin{equation}
v_{\mathrm{Ising}}(c)
=
\sqrt{
r_{+}(1-m)+r_{-}m
}.
\end{equation}

Because the absolute magnitude of this quantity is not fixed by the model, an additional multiplicative scale factor $A_v$ was introduced,
\begin{equation}
v_{\mathrm{model}}(c)
=
A_v\,v_{\mathrm{Ising}}(c).
\end{equation}
For each trial set of $(J_0,\alpha,c_0)$, the optimal value of $A_v$ was calculated analytically by weighted least squares rather than being scanned as an independent nonlinear parameter.

\subsection{Objective function}

The fit was determined by minimizing a combined error function containing contributions from both the mean growth-rate data and the growth-rate heterogeneity data. For the mean growth rate, the normalized residual was
\begin{equation}
R_{g,i}
=
\frac{
g_i-g_{\mathrm{model},i}
}{
\sigma_{g,i}
},
\end{equation}
where $g_i$ is the experimentally measured mean growth rate and $\sigma_{g,i}$ is the experimental uncertainty assigned to the growth-rate measurement.

Each residual was additionally weighted by the magnitude of the measured growth rate,
\begin{equation}
w_{g,i}
=
\frac{|g_i|}
{\left\langle |g| \right\rangle}.
\end{equation}

The growth-rate contribution to the fitting error was therefore
\begin{equation}
\chi_g
=
\sum_i
w_{g,i}R_{g,i}^2.
\end{equation}

For the growth-rate heterogeneity, the normalized residual was
\begin{equation}
R_{v,i}
=
\frac{
v_i-v_{\mathrm{model},i}
}{
\sigma_{v,i}
},
\end{equation}
where $v_i$ is the measured growth-rate heterogeneity and $\sigma_{v,i}$ is its experimental uncertainty.

The weights were determined from the measured heterogeneity values,
\begin{equation}
w_{v,i}
=
\frac{|v_i|}
{\left\langle |v| \right\rangle}.
\end{equation}

The heterogeneity contribution to the fitting error was then
\begin{equation}
\chi_v
=
\sum_i
w_{v,i}R_{v,i}^2.
\end{equation}

As for the case of fitting the Hill function (Eq.\ref{chi_hill}) the total objective function minimized by the fit was
\begin{equation}
\chi_{\mathrm{tot}}
=
W_g\chi_g
+
W_v\chi_v,
\end{equation}
with
\begin{equation}
W_g=1,
\qquad
W_v=0.
\end{equation}

\subsection{Parameter scan}

An initial three-dimensional parameter scan was carried out over $J_0$, $\alpha$, and $c_0$.

The interaction parameter $J_0$ was scanned linearly over the range
\begin{equation}
0.3 \leq J_0 \leq 8.0.
\end{equation}
A total of 49 equally spaced values were used, corresponding to a spacing
\begin{equation}
\Delta J_0
=
\frac{8.0-0.3}{49-1}
\simeq 0.1604.
\end{equation}

The parameter $\alpha$ was scanned logarithmically using 35 points. Because the experimental concentration range differs between antibiotics, the $\alpha$ range was defined separately for each antibiotic. If $c_{\min}$ and $c_{\max}$ denote the smallest non-zero and largest experimental concentrations, respectively, the scan range was
\begin{equation}
\alpha_{\min}
=
\frac{0.1}{c_{\max}},
\end{equation}
and
\begin{equation}
\alpha_{\max}
=
\frac{1000}{c_{\min}}.
\end{equation}

Thus, for each antibiotic, 35 logarithmically spaced values of $\alpha$ were evaluated between these limits.

The concentration offset $c_0$ was also scanned logarithmically using 35 points. Its range was defined separately for each antibiotic as
\begin{equation}
c_{0,\min}
=
10^{-4}c_{\min},
\end{equation}
and
\begin{equation}
c_{0,\max}
=
c_{\max}.
\end{equation}

Consequently, the initial coarse search evaluated
\begin{equation}
49 \times 35 \times 35
=
60025
\end{equation}
combinations of $(J_0,\alpha,c_0)$ for each antibiotic.

For each parameter combination, the Ising model was solved numerically at all experimentally measured concentrations, the optimal heterogeneity scale factor $A_v$ was determined analytically, and the total objective function $\chi_{\mathrm{tot}}$ was calculated. The parameter combination giving the smallest value of $\chi_{\mathrm{tot}}$ was retained as the best result of the coarse scan.

\subsection{Continuous parameter refinement}

The best parameter combination obtained from the coarse scan was subsequently used as the starting point for a continuous local optimization using MATLAB's \texttt{fminsearch} routine.

During this refinement, $J_0$ was optimized directly, while $\alpha$ and $c_0$ were optimized in logarithmic coordinates,
\begin{equation}
p_1 = J_0,
\qquad
p_2 = \ln\alpha,
\qquad
p_3 = \ln c_0.
\end{equation}
The logarithmic representation guarantees that $\alpha$ and $c_0$ remain positive.

The refinement was constrained to remain inside the parameter ranges explored by the coarse scan. Trial parameter sets outside these limits were assigned a large penalty in the objective function. At every step of the continuous optimization, the Ising model was recalculated, the optimal heterogeneity scale factor $A_v$ was recalculated analytically, and the complete weighted objective function was reevaluated.

The final reported parameters $(J_0,\alpha,c_0)$ were therefore not restricted to the discrete values of the original scan. The coarse scan was used to locate the appropriate region of parameter space, while the subsequent continuous optimization provided the final refined parameter estimates.

\section{Comparison of Ising model fits with different error function weights}

To account for local variations in the measured growth rate, the fitting error was weighted using the growth-rate values at each concentration and its two nearest neighboring concentrations. Specifically, for an interior data point $i$, we defined the local average growth rate as
\begin{equation}
\widetilde{g}_i =
\frac{|g_{i-1}|+|g_i|+|g_{i+1}|}{3},
\end{equation}
where $g_i$ is the experimentally measured mean growth rate at concentration $c_i$. For the first and last data points, the average was calculated using only the point itself and its single available neighbor.

The normalized weight assigned to each data point was then
\begin{equation}
w_i = \frac{\widetilde{g}_i}
{\langle \widetilde{g} \rangle},
\end{equation}
where $\langle \widetilde{g} \rangle$ denotes the average of the local growth-rate values over all measured concentrations.

The weighted fitting error for the mean growth rate was calculated as
\begin{equation}
\chi_g =
\sum_i w_i
\left(
\frac{g_i-g_{\mathrm{model}}(c_i)}
{\sigma_{g,i}}
\right)^2,
\end{equation}
where $\sigma_{g,i}$ is the experimental uncertainty in the growth-rate measurement. Importantly, the local averaging was applied only to the weights, while the fitting residuals were calculated using the original, unsmoothed experimental measurements.

\begin{figure}[htbp]
    \centering

    % Panel (a)
    \begin{subfigure}{0.8\linewidth}
        \centering
        \includegraphics[width=\linewidth]{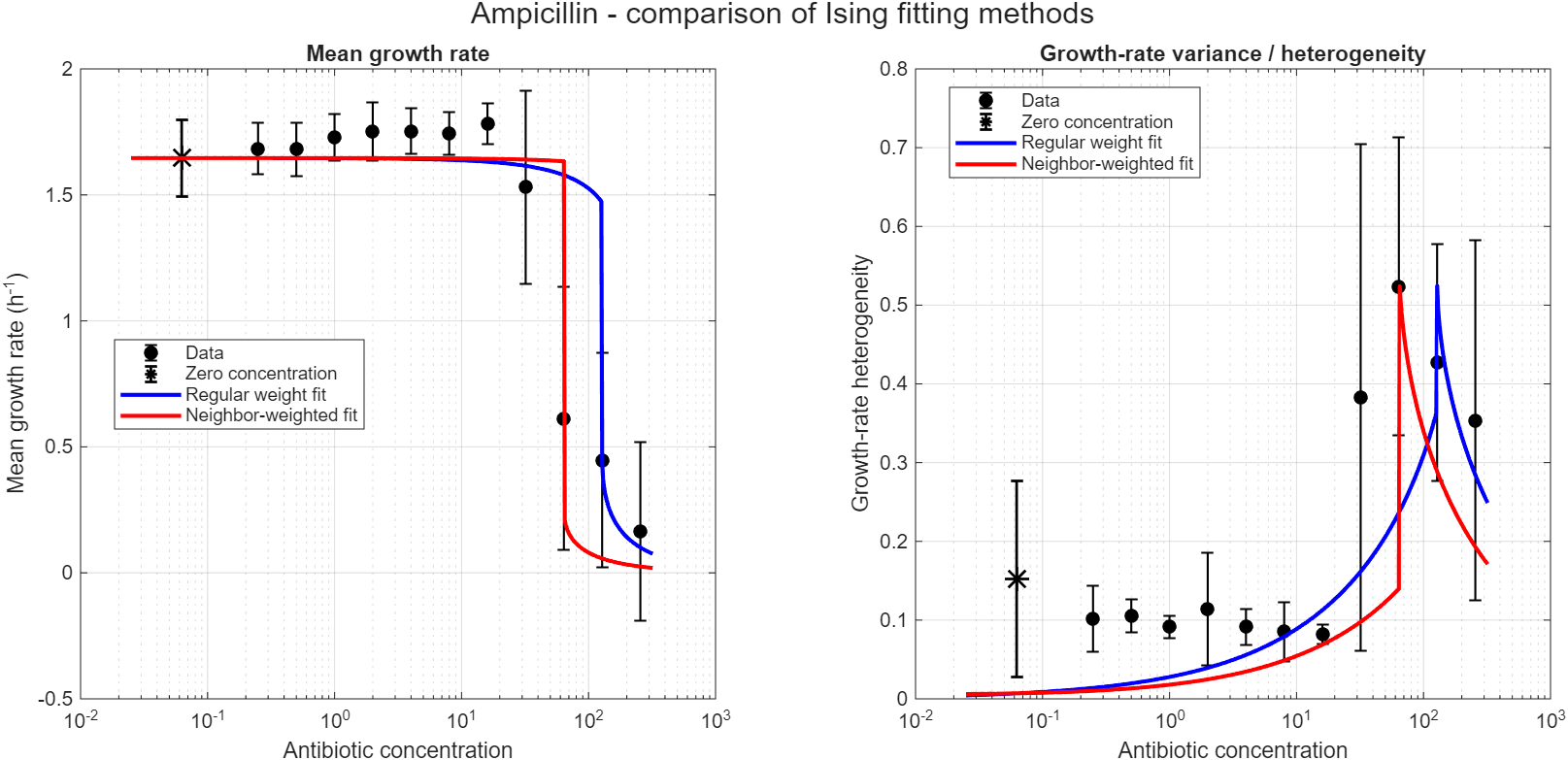}
        \caption{}
        \label{fig:panel_a}
    \end{subfigure}

    \vspace{0.1cm}

    % Panel (b)
    \begin{subfigure}{0.8\linewidth}
        \centering
        \includegraphics[width=\linewidth]{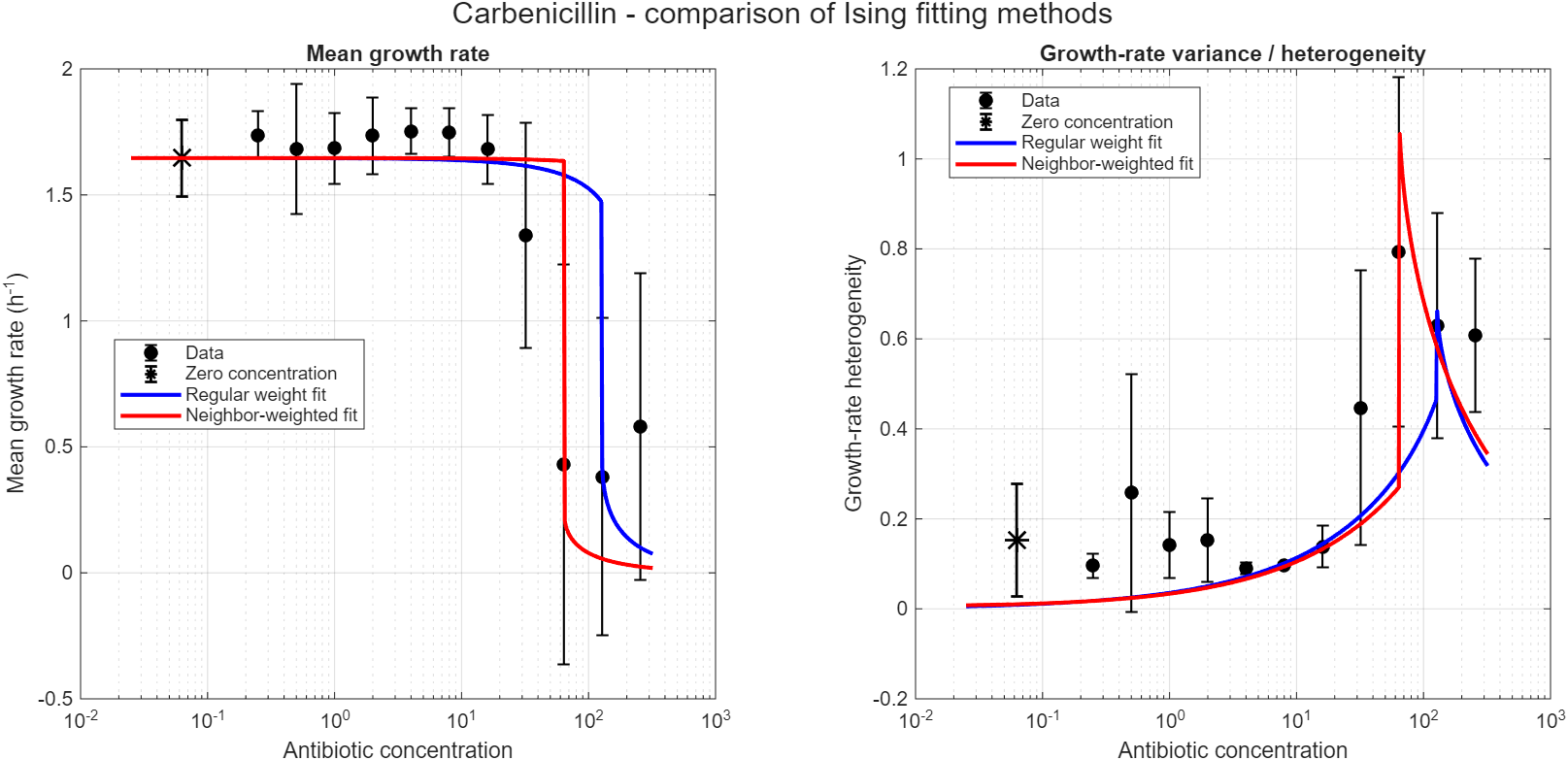}
        \caption{}
        \label{fig:panel_b}
    \end{subfigure}

    \vspace{0.1cm}

    % Panel (c)
    \begin{subfigure}{0.8\linewidth}
        \centering
        \includegraphics[width=\linewidth]{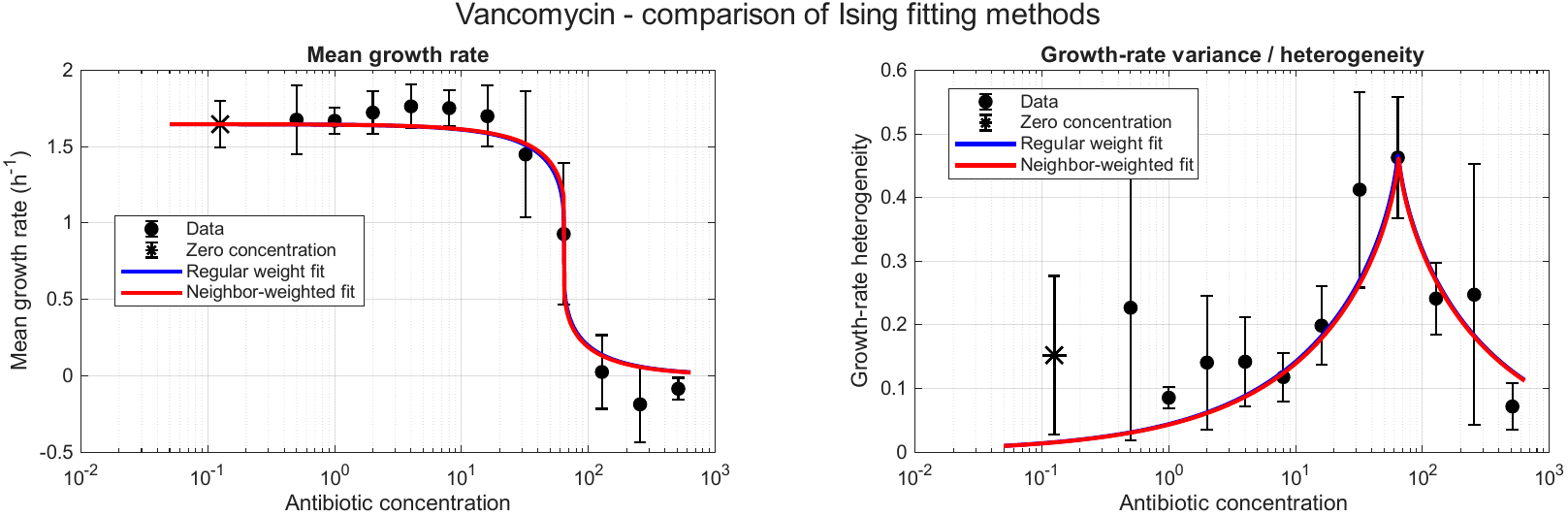}
        \caption{}
        \label{fig:panel_c}
    \end{subfigure}
\captionsetup{justification=raggedright,singlelinecheck=false}
    \caption{Comparison of experimental data (for \emph{E. coli}) and Ising model fits, demonstrating the effect of using in the error calculation weights averaged over three neighboring points of the growth-rate data. The main effect is to move the sharp jump in the Ising model curve to lower antibiotic concentrations, which seems more reasonable for the cases of (a) Ampicillin and (b) Carbenicillin. For all the other cases, as shown for example for (c) Vancomycin, the effect is negligible. The fitting parameters, are shown in Table \ref{tab:ising_parameters_comparison}.}

    \label{fig:three_panels}
\end{figure}

While this procedure reduces the sensitivity of the weighting to fluctuations at individual data points, it was most important when fitting the Ising model to a few data sets due to this method giving greater importance to measurements in concentration regions characterized by higher growth rates. As can be seen in Fig.\ref{fig:three_panels}\subref{fig:panel_a},\subref{fig:panel_b} this improved the fitting by shifting the sharp jump in the Ising model graph to lower antibiotic concentrations. For all other cases, both methods give very similar fitting parameters, as shown in Table \ref{tab:ising_parameters_comparison}.

\begin{table}[htbp]
\centering
\captionsetup{justification=raggedright,singlelinecheck=false}
\caption{Comparison of the fitted Ising-model parameters obtained
using the regular growth-rate weighting and the three-point
neighbor-weighted fitting procedure (Fig.\ref{fig:three_panels}).
$J_0$ is the interaction strength, $\alpha$ is the
concentration-scaling parameter, $c_0$ is the concentration
offset, and $A_v$ is the multiplicative scaling factor
for the predicted growth-rate heterogeneity.
In both procedures, only the growth-rate fitting error
was minimized ($W_g=1$, $W_v=0$).}
\label{tab:ising_parameters_comparison}

\renewcommand{\arraystretch}{1.3}

\begin{tabular}{|l|l|c|c|c|c|}
\hline
\textbf{Antibiotic} &
\textbf{Fitting method} &
$\mathbf{J_0}$ &
$\boldsymbol{\alpha}$ &
$\mathbf{c_0}$ &
$\mathbf{A_v}$ \\
\hline

Ampicillin
& Regular
& 5.000
& 4.448
& $4.137\times10^{-5}$
& 0.8392 \\
\hline

Ampicillin
& Neighbor-weighted
& 7.795
& 15.538
& 0.1065
& 1.1151 \\
\hline

Carbenicillin
& Regular
& 5.000
& 4.457
& $2.500\times10^{-5}$
& 1.0728 \\
\hline

Carbenicillin
& Neighbor-weighted
& 7.908
& 15.820
& 0.03422
& 2.2631 \\
\hline

Vancomycin
& Regular
& 4.090
& 6.564
& $5.000\times10^{-5}$
& 0.6924 \\
\hline

Vancomycin
& Neighbor-weighted
& 4.196
& 6.855
& $1.651\times10^{-4}$
& 0.7005 \\
\hline

\end{tabular}
\end{table}

\section{Comparison of fits of the Hills and Ising functions to the experimental data of \emph{E. coli}.}

\begin{figure}[p]
    \centering

    \includegraphics[
        width=\textwidth,
        height=0.85\textheight,
        keepaspectratio
    ]{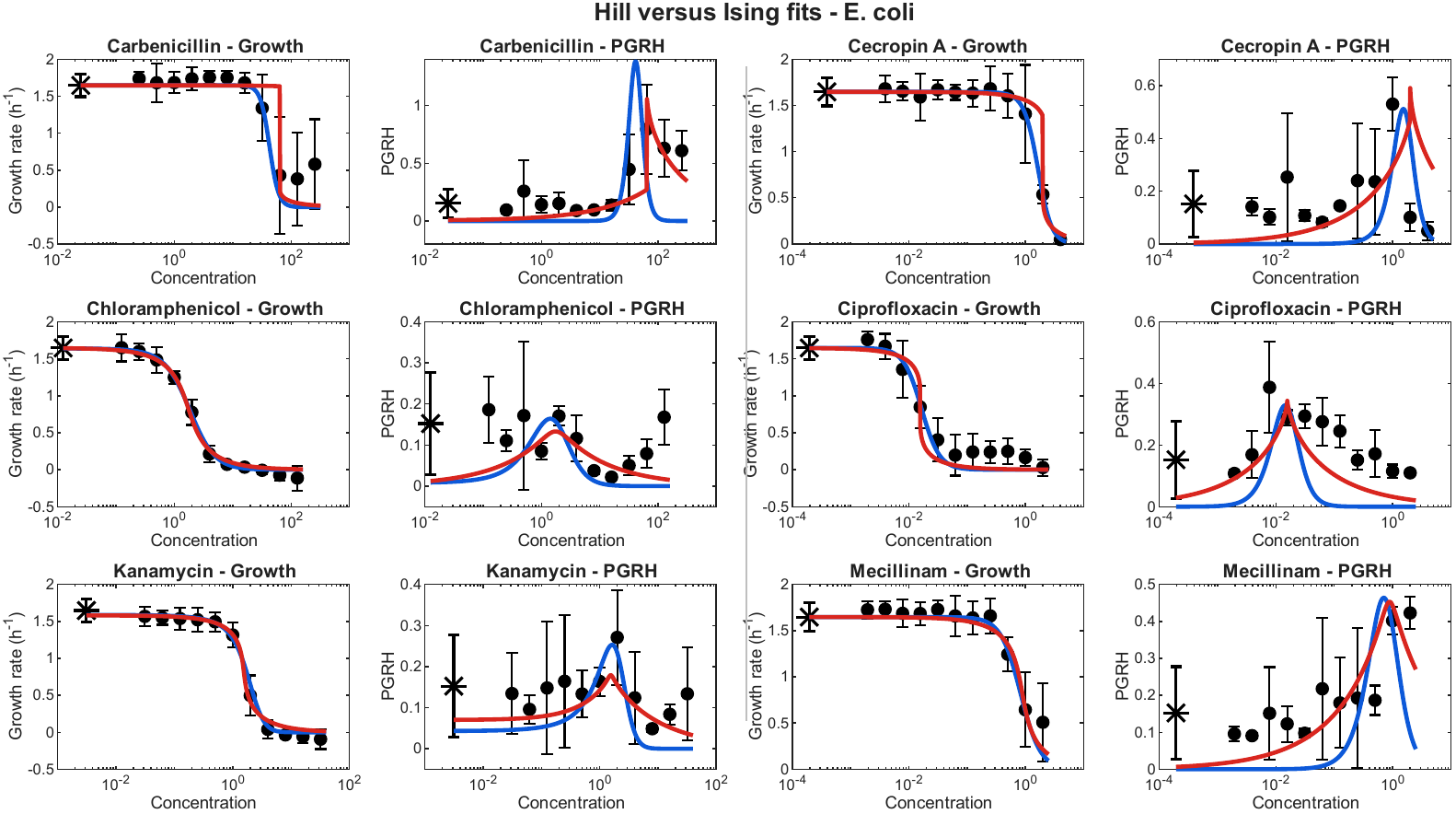}

    \caption{
    \textbf{Comparison of Hill and Ising models fits to bacterial growth
    rates and growth-rate heterogeneity, for \emph{E. coli}.}
    Each antibiotic is represented by two panels:
    mean growth rate (left) and PGRH (right). Black symbols represent the experimental measurements,
    with error bars indicating the experimental uncertainties.
    Blue curves show fits using Hill model, while red curves show the Ising model fits.
    }

    \label{fig:HillIsingSupplementary1}
\end{figure}

\begin{figure}[p]
    \centering

    \includegraphics[
        width=\textwidth,
        height=0.85\textheight,
        keepaspectratio
    ]{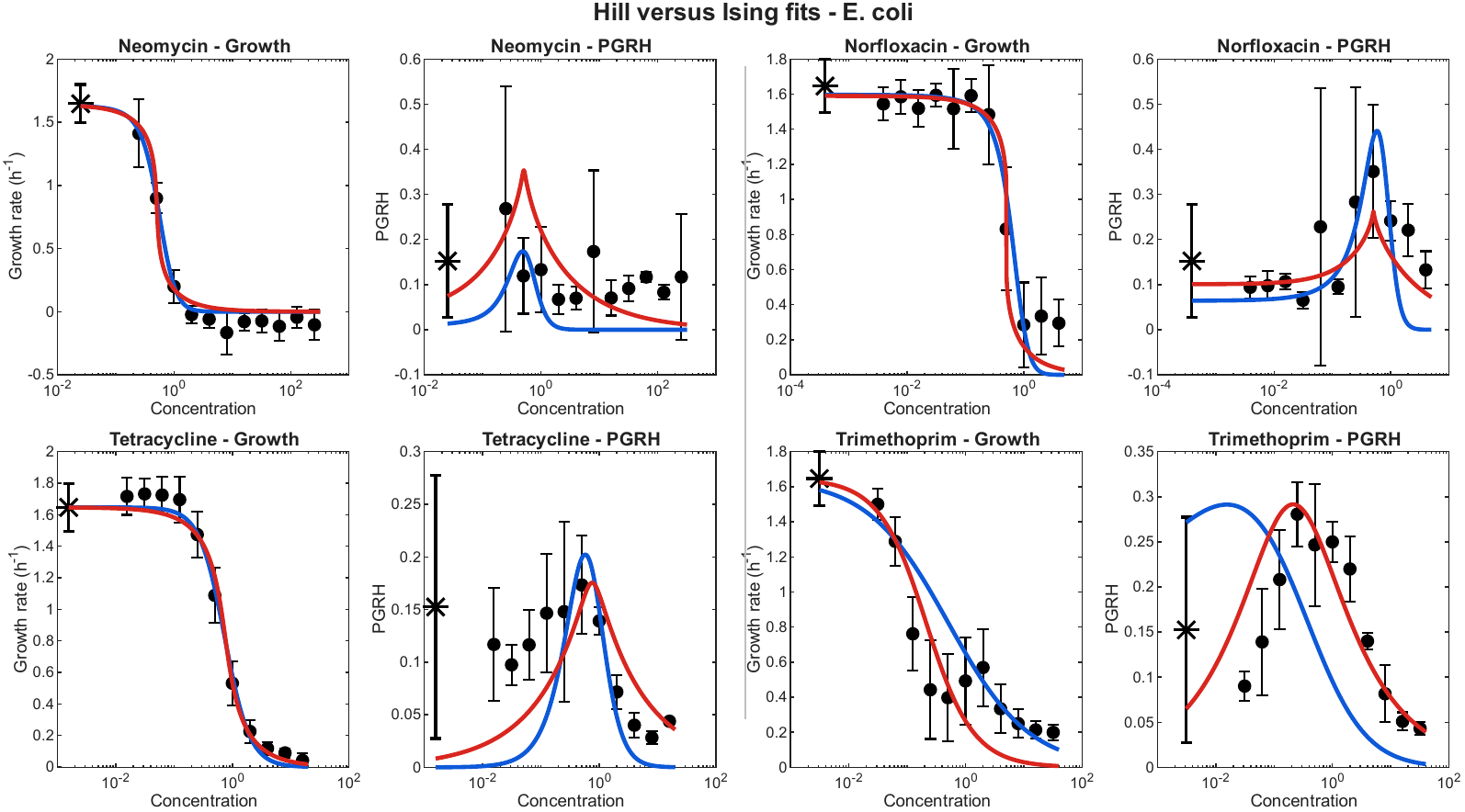}

    \caption{
    \textbf{Comparison of Hill and Ising models fits to bacterial growth
    rates and growth-rate heterogeneity, for \emph{E. coli}.}
    The organization and symbols are the same as in
    Fig.~\ref{fig:HillIsingSupplementary1}.
    Each antibiotic is represented by two panels:
    mean growth rate (left) and PGRH (right).
    Blue curves show fits using Hill model, while red curves show the Ising model fits.
    }

    \label{fig:HillIsingSupplementary2}
\end{figure}

In Figs.\ref{fig:HillIsingSupplementary1},\ref{fig:HillIsingSupplementary2} we give the comparisons of the Hill and Ising models to the experimental data, for the different antibiotics not shown in Fig.1. The fitting parameters for all the antibiotics are given in Table \ref{tab:HillParameters} and Table \ref{tab:IsingParameters}, for the Hill and Ising models respectively.

\begin{table}[htbp]
\centering
\caption{Fitted parameters of the Hill model for the different
antibiotics. Here $c_H$ is the characteristic Hill concentration,
$n_H$ is the Hill exponent, $c_0$ is the concentration offset,
and $A_h$ is the multiplicative factor used to fit the
growth-rate heterogeneity (PGRH).
The fitting was performed using $W_g=1$ and $W_v=0$.}
\label{tab:HillParameters}

\renewcommand{\arraystretch}{1.2}

\begin{tabular}{|l|c|c|c|c|}
\hline
\textbf{Antibiotic} &
$\boldsymbol{c_H}$ &
$\boldsymbol{n_H}$ &
$\boldsymbol{c_0}$ &
$\boldsymbol{A_h}$ \\
\hline
Ampicillin       & 58.9365 & 6.0298  & $2.5\times10^{-5}$   & 1.91654 \\
\hline
Carbenicillin    & 42.7270 & 5.61533 & $2.5\times10^{-5}$   & 3.87670 \\
\hline
Cecropin A       & 1.79743 & 4.02253 & 0.157656            & 0.407625 \\
\hline
Chloramphenicol  & 1.94266 & 2.12842 & 0.131298            & 0.246893 \\
\hline
Ciprofloxacin    & 0.0165462 & 2.81359 & $1.95\times10^{-7}$ & 0.0354153 \\
\hline
Gentamicin       & 0.329107 & 2.14751 & 0.0262875          & 0.149701 \\
\hline
Kanamycin        & 5.10063 & 8.00000 & 3.38286             & 0.172658 \\
\hline
Mecillinam       & 0.826947 & 2.54980 & $1.95\times10^{-7}$ & 0.386467 \\
\hline
Neomycin         & 0.762646 & 4.06429 & 0.230378           & 0.0893916 \\
\hline
Norfloxacin      & 1.73493 & 8.00000 & 1.12542             & 0.175571 \\
\hline
Rifampicin       & 11.1439 & 8.00000 & 4.04953             & 0.373598 \\
\hline
Tetracycline     & 0.738657 & 2.01476 & $1.56\times10^{-6}$ & 0.196839 \\
\hline
Trimethoprim     & 0.509826 & 0.626311 & $3.125\times10^{-6}$ & 0.387788 \\
\hline
Vancomycin       & 66.8485 & 3.80755 & $5.0\times10^{-5}$   & 2.75269 \\
\hline
\end{tabular}
\end{table}

\begin{table}[htbp]
\centering
\caption{Fitted parameters of the Ising model for the different
antibiotics. Here $J_0$ is the effective interaction strength,
$\alpha$ is the concentration-scaling parameter, $c_0$ is
the concentration offset, and $A_i$ is the multiplicative
factor used to fit the growth-rate heterogeneity (PGRH).
The growth-rate fitting error was weighted using the local
average of three neighboring growth-rate measurements.
The fitting was performed using $W_g=1$ and $W_v=0$.}
\label{tab:IsingParameters}

\renewcommand{\arraystretch}{1.2}

\begin{tabular}{|l|c|c|c|c|}
\hline
\textbf{Antibiotic} &
$\boldsymbol{J_0}$ &
$\boldsymbol{\alpha}$ &
$\boldsymbol{c_0}$ &
$\boldsymbol{A_i}$ \\
\hline
Ampicillin       & 7.79507 & 15.5379 & 0.106475 & 1.11514 \\
\hline
Carbenicillin    & 7.90770 & 15.8202 & 0.0342227 & 2.26314 \\
\hline
Cecropin A       & 4.64634 & 256.539 & $3.90625\times10^{-7}$ & 0.910229 \\
\hline
Chloramphenicol  & 2.40967 & 104.224 & $1.25\times10^{-5}$ & 0.188290 \\
\hline
Ciprofloxacin    & 4.14003 & 27440.7 & $2.10563\times10^{-6}$ & 0.508554 \\
\hline
Gentamicin       & 2.49310 & 651.797 & $6.25\times10^{-6}$ & 0.287481 \\
\hline
Kanamycin        & 3.55724 & 164.047 & 0.420246 & 0.253156 \\
\hline
Mecillinam       & 2.93382 & 266.775 & $1.95\times10^{-7}$ & 0.640365 \\
\hline
Neomycin         & 3.66995 & 660.036 & 0.0104346 & 0.499939 \\
\hline
Norfloxacin      & 4.13308 & 645.774 & 0.162365 & 0.391894 \\
\hline
Rifampicin       & 4.68664 & 55.7757 & 1.31499 & 0.357059 \\
\hline
Tetracycline     & 2.57864 & 262.591 & $1.56\times10^{-6}$ & 0.248076 \\
\hline
Trimethoprim     & 0.30000 & 298.543 & $3.125\times10^{-6}$ & 0.412273 \\
\hline
Vancomycin       & 4.19631 & 6.85493 & $1.65068\times10^{-4}$ & 0.700502 \\
\hline
\end{tabular}
\end{table}

\section{Comparison of fits of the Hills and Ising functions to the experimental data of \emph{Staph}.}

In Figs.\ref{fig:HillIsingSupplementary1},\ref{fig:HillIsingSupplementary2} we give the comparisons of the Hill and Ising models to the experimental data, for the different antibiotics not shown in Fig.1. The fitting parameters for all the antibiotics are given in Table \ref{tab:HillParametersStaph} and Table \ref{tab:IsingParametersStaph}, for the Hill and Ising models respectively.

\begin{figure}[p]
    \centering

    \includegraphics[
        width=\textwidth,
        height=0.85\textheight,
        keepaspectratio
    ]{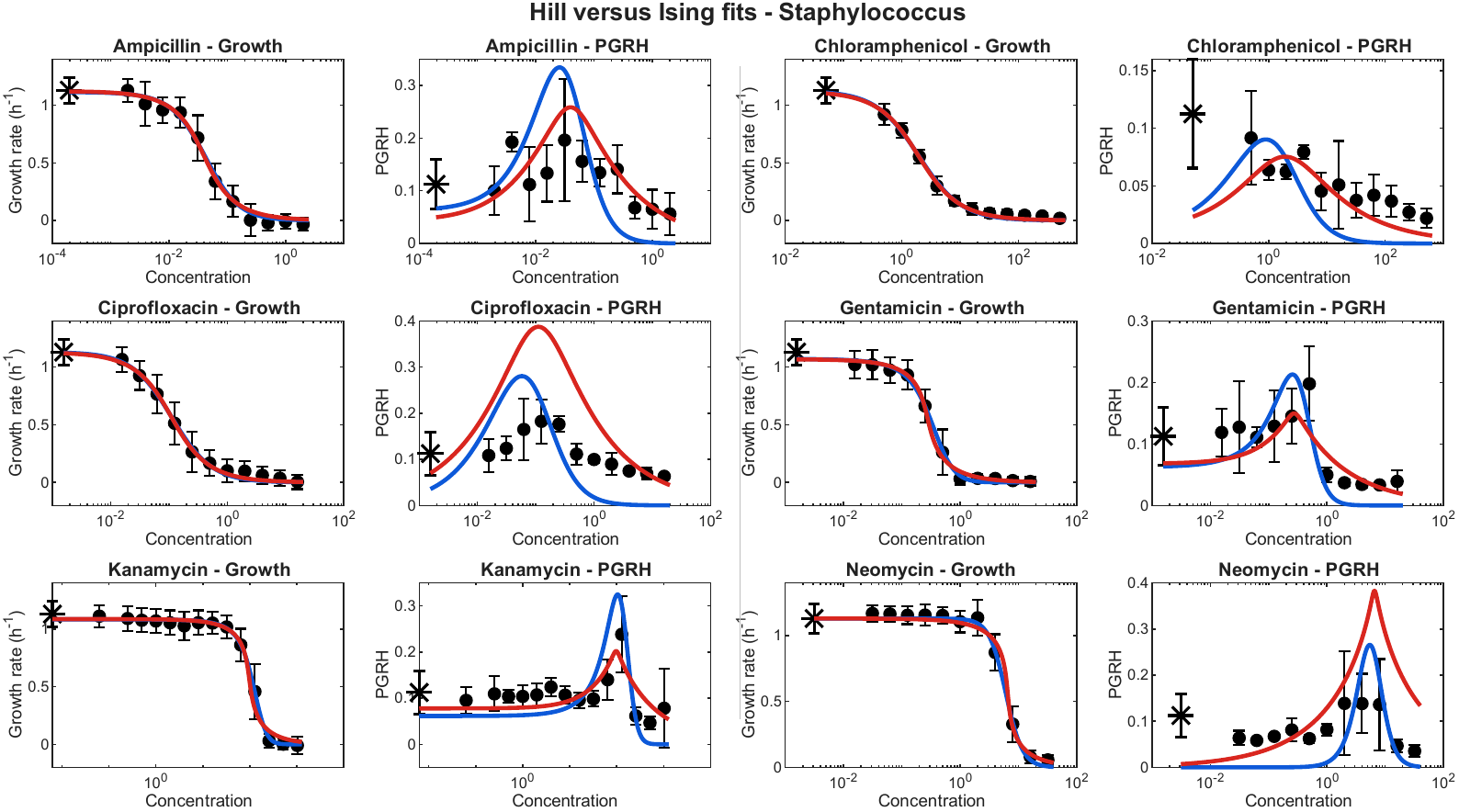}

    \caption{
    \textbf{Comparison of Hill and Ising models fits to bacterial growth
    rates and growth-rate heterogeneity, for \emph{Staph}.}
    Each antibiotic is represented by two panels:
    mean growth rate (left) and PGRH (right). Black symbols represent the experimental measurements,
    with error bars indicating the experimental uncertainties.
    Blue curves show fits using Hill model, while red curves show the Ising model fits.
    }

    \label{fig:HillIsingSupplementary1staph}
\end{figure}

\begin{figure}[p]
    \centering

    \includegraphics[
        width=\textwidth,
        height=0.85\textheight,
        keepaspectratio
    ]{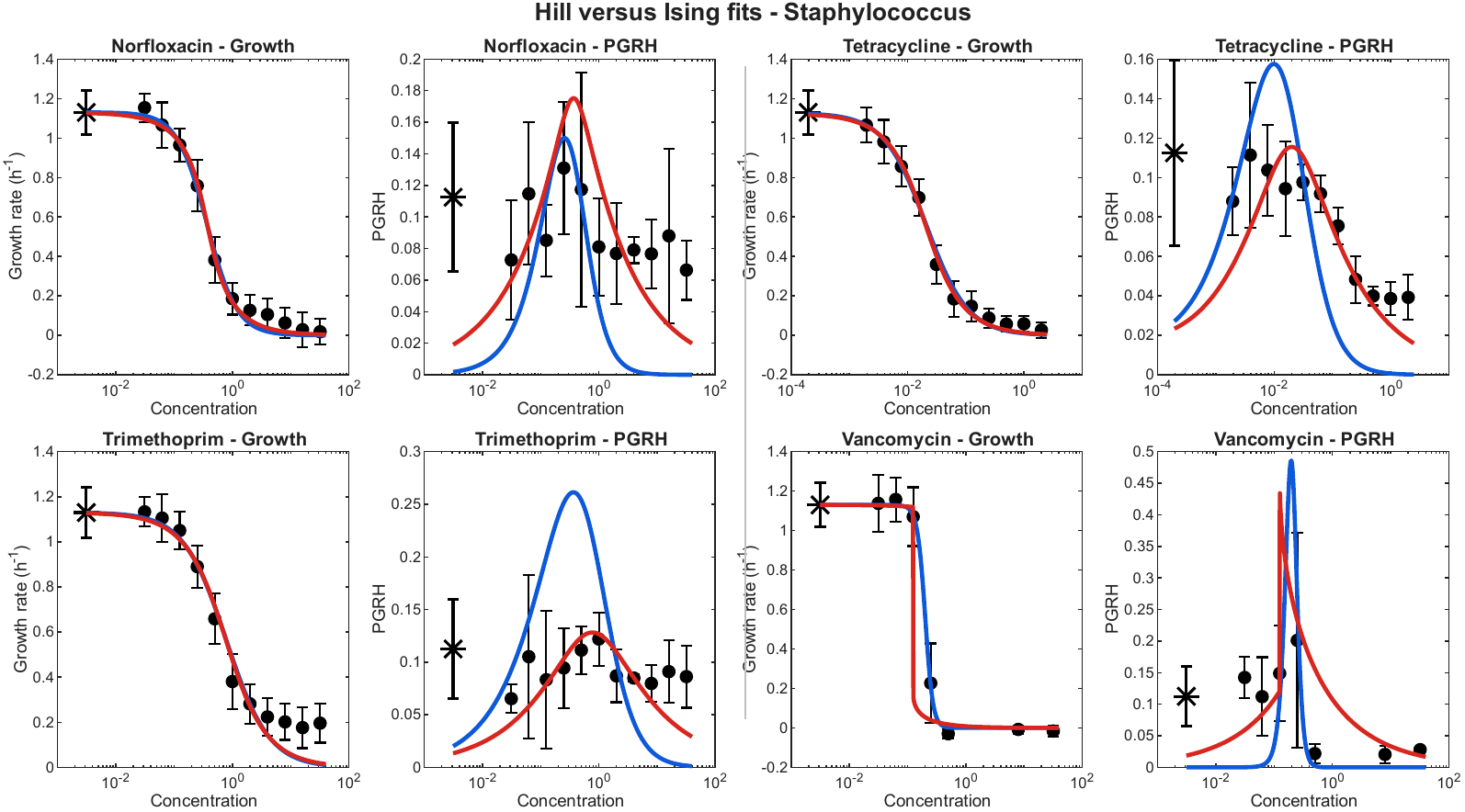}

    \caption{
    \textbf{Comparison of Hill and Ising models fits to bacterial growth
    rates and growth-rate heterogeneity, for \emph{Staph}.}
    Each antibiotic is represented by two panels:
    mean growth rate (left) and PGRH (right). Black symbols represent the experimental measurements,
    with error bars indicating the experimental uncertainties.
    Blue curves show fits using Hill model, while red curves show the Ising model fits.
    }

    \label{fig:HillIsingSupplementary2staph}
\end{figure}

\begin{table}[htbp]
\centering

\caption{
Fitted parameters of the Hill model for antibiotics
acting on Staphylococcus.
Here $c_H$ is the characteristic Hill concentration,
$n_H$ is the Hill exponent, $c_0$ is the concentration
offset, and $A_h$ is the multiplicative factor used
to fit the population growth-rate heterogeneity (PGRH).
The fitting was performed with $W_g=1$ and $W_v=0$.
}

\label{tab:HillParametersStaph}

\renewcommand{\arraystretch}{1.2}

\begin{tabular}{|l|c|c|c|c|}
\hline
\textbf{Antibiotic} &
$\boldsymbol{c_H}$ &
$\boldsymbol{n_H}$ &
$\boldsymbol{c_0}$ &
$\boldsymbol{A_h}$ \\
\hline

Ampicillin
& 0.0433942 & 1.55724 & 0.0026907 & 0.142502 \\
\hline

Chloramphenicol
& 1.93786 & 1.18903 & 0.0179395 & 0.311659 \\
\hline

Ciprofloxacin
& 0.111616 & 1.26832 & $1.56\times10^{-6}$ & 0.222896 \\
\hline

Gentamicin
& 0.523863 & 3.43106 & 0.224161 & 0.155704 \\
\hline

Kanamycin
& 337.842 & 8.00000 & 227.494 & 2.63222 \\
\hline

Neomycin
& 6.07828 & 3.02801 & $3.125\times10^{-6}$ & 0.744778 \\
\hline

Norfloxacin
& 0.367182 & 1.69730 & $3.125\times10^{-6}$ & 0.173554 \\
\hline

Tetracycline
& 0.0207333 & 1.21225 & $1.31161\times10^{-4}$ & 0.0555851 \\
\hline

Trimethoprim
& 0.786180 & 1.18100 & $3.125\times10^{-6}$ & 0.577211 \\
\hline

Vancomycin
& 0.232749 & 7.54682 & 0.0321128 & 0.109480 \\
\hline

\end{tabular}
\end{table}

\begin{table}[htbp]
\centering

\caption{
Fitted parameters of the Ising model for antibiotics
acting on Staphylococcus.
Here $J_0$ is the effective interaction strength,
$\alpha$ is the concentration-scaling parameter,
$c_0$ is the concentration offset, and $A_i$ is
the multiplicative factor used to fit PGRH.
The growth-rate fitting error was weighted using
the local average of three neighboring growth-rate
measurements. The fitting was performed with
$W_g=1$ and $W_v=0$.
}

\label{tab:IsingParametersStaph}

\renewcommand{\arraystretch}{1.2}

\begin{tabular}{|l|c|c|c|c|}
\hline
\textbf{Antibiotic} &
$\boldsymbol{J_0}$ &
$\boldsymbol{\alpha}$ &
$\boldsymbol{c_0}$ &
$\boldsymbol{A_i}$ \\
\hline

Ampicillin
& 1.59985 & 3024.97 & $6.37185\times10^{-4}$ & 0.365473 \\
\hline

Chloramphenicol
& 0.797820 & 42.2574 & 0.0201506 & 0.106536 \\
\hline

Ciprofloxacin
& 1.07382 & 842.597 & $1.56\times10^{-6}$ & 0.547975 \\
\hline

Gentamicin
& 3.05664 & 713.552 & 0.0790621 & 0.211718 \\
\hline

Kanamycin
& 3.31935 & 2.40182 & 22.1676 & 0.285990 \\
\hline

Neomycin
& 3.45607 & 46.9218 & $3.125\times10^{-6}$ & 0.541882 \\
\hline

Norfloxacin
& 2.10603 & 424.569 & $3.125\times10^{-6}$ & 0.247741 \\
\hline

Tetracycline
& 1.05533 & 4564.11 & $1.60884\times10^{-4}$ & 0.163298 \\
\hline

Trimethoprim
& 0.782754 & 104.075 & $3.125\times10^{-6}$ & 0.181277 \\
\hline

Vancomycin
& 7.62059 & 7758.26 & $3.13025\times10^{-6}$ & 0.890342 \\
\hline

\end{tabular}
\end{table}

\section{Comparison of fits of the Hills and Ising functions to the experimental data of \emph{Pseudomonas}.}

\begin{figure}[p]
    \centering

    \includegraphics[
        width=\textwidth,
        height=0.85\textheight,
        keepaspectratio
    ]{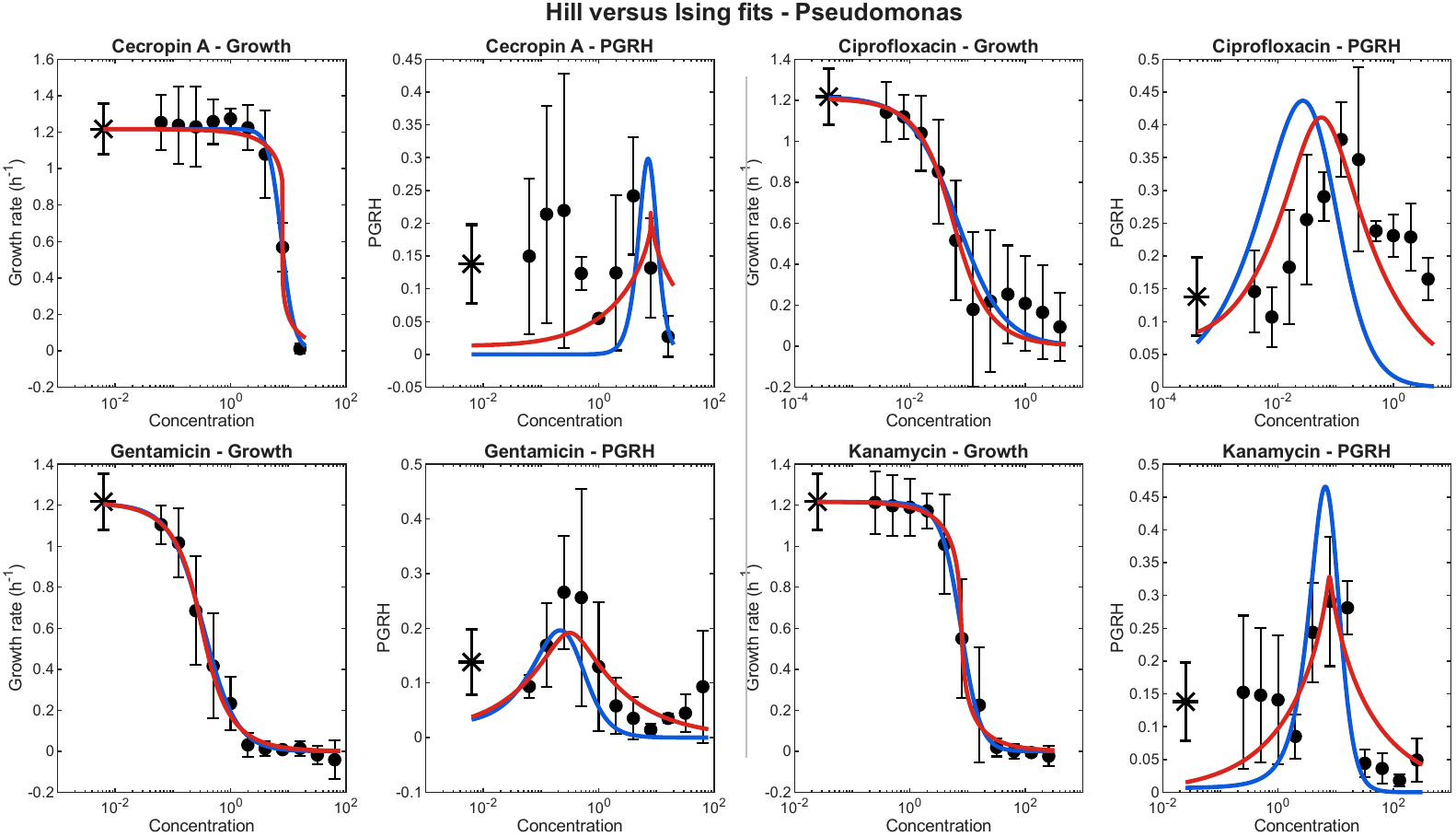}

    \caption{
    \textbf{Comparison of Hill and Ising models fits to bacterial growth
    rates and growth-rate heterogeneity, for \emph{Pseu}.}
    Each antibiotic is represented by two panels:
    mean growth rate (left) and PGRH (right). Black symbols represent the experimental measurements,
    with error bars indicating the experimental uncertainties.
    Blue curves show fits using Hill model, while red curves show the Ising model fits.
    }

    \label{fig:HillIsingSupplementary1pseu}
\end{figure}

\begin{figure}[p]
    \centering

    \includegraphics[
        width=\textwidth,
        height=0.85\textheight,
        keepaspectratio
    ]{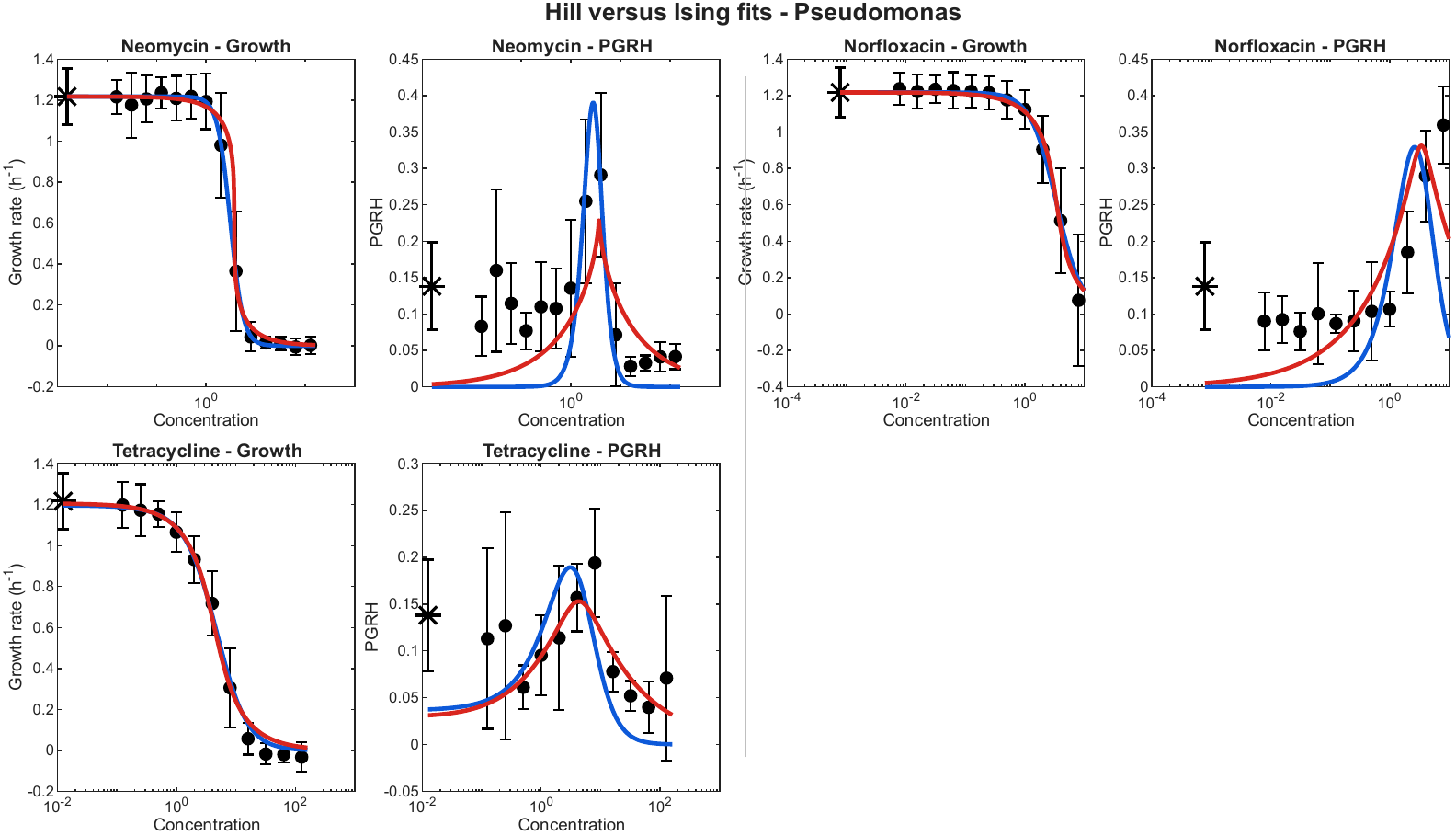}

    \caption{
    \textbf{Comparison of Hill and Ising models fits to bacterial growth
    rates and growth-rate heterogeneity, for \emph{Pseu}.}
    Each antibiotic is represented by two panels:
    mean growth rate (left) and PGRH (right). Black symbols represent the experimental measurements,
    with error bars indicating the experimental uncertainties.
    Blue curves show fits using Hill model, while red curves show the Ising model fits.
    }

    \label{fig:HillIsingSupplementary2pseu}
\end{figure}

In Figs.\ref{fig:HillIsingSupplementary1pseu},\ref{fig:HillIsingSupplementary2pseu} we give the comparisons of the Hill and Ising models to the experimental data, for the different antibiotics not shown in Fig.1. The fitting parameters for all the antibiotics are given in Table \ref{tab:HillParametersPseu} and Table \ref{tab:IsingParametersPseu}, for the Hill and Ising models respectively.

\begin{table}[htbp]
\centering

\caption{
Fitted parameters of the Hill model for antibiotics
acting on Pseudomonas.
Here $c_H$ is the characteristic Hill concentration,
$n_H$ is the Hill exponent, $c_0$ is the concentration
offset, and $A_h$ is the multiplicative factor used
to fit the population growth-rate heterogeneity (PGRH).
The fitting was performed with $W_g=1$ and $W_v=0$.
}

\label{tab:HillParametersPseu}

\renewcommand{\arraystretch}{1.2}

\begin{tabular}{|l|c|c|c|c|}
\hline
\textbf{Antibiotic} &
$\boldsymbol{c_H}$ &
$\boldsymbol{n_H}$ &
$\boldsymbol{c_0}$ &
$\boldsymbol{A_h}$ \\
\hline

Cecropin A
& 7.66962 & 4.00720 & $6.25\times10^{-6}$ & 0.666977 \\
\hline

Ciprofloxacin
& 0.0671101 & 1.08862 & $3.90625\times10^{-7}$ & 0.274307 \\
\hline

Gentamicin
& 0.348592 & 1.55534 & 0.0132142 & 0.220277 \\
\hline

Kanamycin
& 8.34651 & 2.81205 & 0.722343 & 1.52324 \\
\hline

Neomycin
& 3.07602 & 3.35123 & $1.5625\times10^{-6}$ & 0.656247 \\
\hline

Norfloxacin
& 3.40274 & 1.98838 & $7.8125\times10^{-7}$ & 0.941797 \\
\hline

Tetracycline
& 4.97791 & 1.77014 & 0.499027 & 0.723106 \\
\hline

\end{tabular}
\end{table}
\begin{table}[htbp]
\centering

\caption{
Fitted parameters of the Ising model for antibiotics
acting on Pseudomonas.
Here $J_0$ is the effective interaction strength,
$\alpha$ is the concentration-scaling parameter,
$c_0$ is the concentration offset, and $A_i$ is
the multiplicative factor used to fit PGRH.
The growth-rate fitting error was weighted using
the local average of three neighboring growth-rate
measurements. The fitting was performed with
$W_g=1$ and $W_v=0$.
}

\label{tab:IsingParametersPseu}

\renewcommand{\arraystretch}{1.2}

\begin{tabular}{|l|c|c|c|c|}
\hline
\textbf{Antibiotic} &
$\boldsymbol{J_0}$ &
$\boldsymbol{\alpha}$ &
$\boldsymbol{c_0}$ &
$\boldsymbol{A_i}$ \\
\hline

Cecropin A
& 4.30979 & 56.8104 & 0.0612377 & 0.317610 \\
\hline

Ciprofloxacin
& 1.19797 & 1739.23 & $6.97631\times10^{-4}$ & 0.581538 \\
\hline

Gentamicin
& 1.79765 & 424.222 & 0.00130184 & 0.270968 \\
\hline

Kanamycin
& 3.60624 & 42.3346 & $2.5\times10^{-5}$ & 0.464707 \\
\hline

Neomycin
& 4.03252 & 111.301 & $1.5625\times10^{-6}$ & 0.331152 \\
\hline

Norfloxacin
& 2.70176 & 61.7470 & $7.8125\times10^{-7}$ & 0.468227 \\
\hline

Tetracycline
& 1.88093 & 31.7356 & 0.104785 & 0.216291 \\
\hline

\end{tabular}
\end{table}

\section{Comparison of the Ising interaction parameter $J_0$ and Hill exponent $n_{hill}$ to the experimental data of the growth-rate.}

\begin{figure}[htbp]
    \centering

    % Panel (a)
    \begin{subfigure}{0.8\linewidth}
        \centering
        \includegraphics[width=0.7\linewidth]{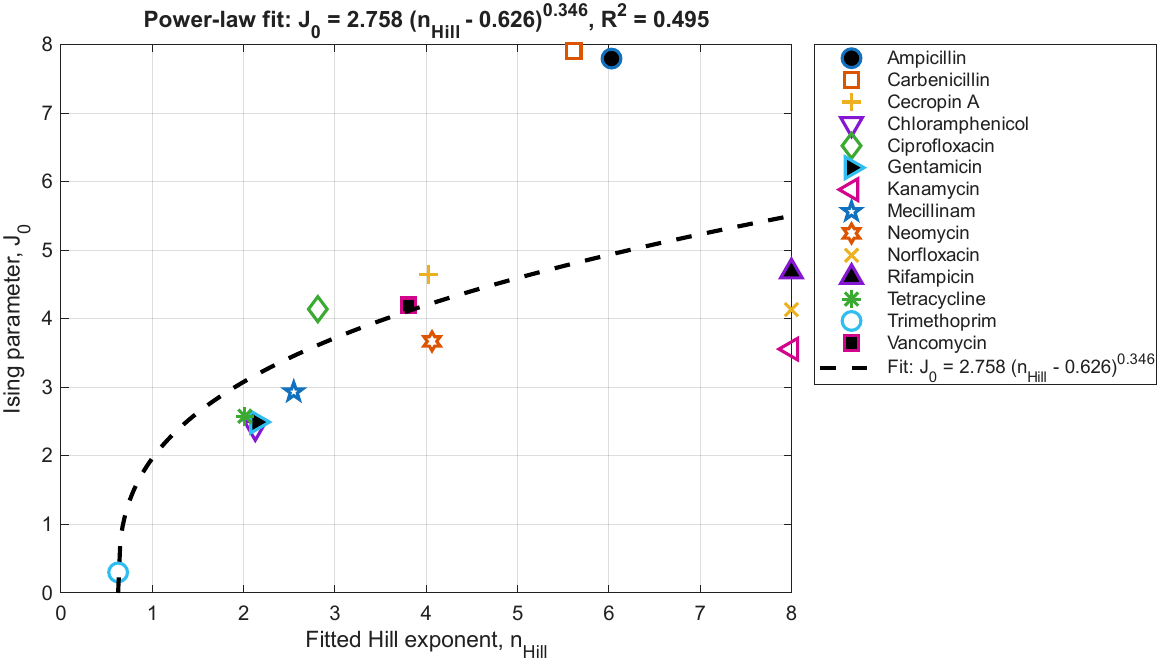}
        \caption{}
        \label{fig:panel_a}
    \end{subfigure}

    \vspace{0.1cm}

    % Panel (b)
    \begin{subfigure}{0.8\linewidth}
        \centering
        \includegraphics[width=0.7\linewidth]{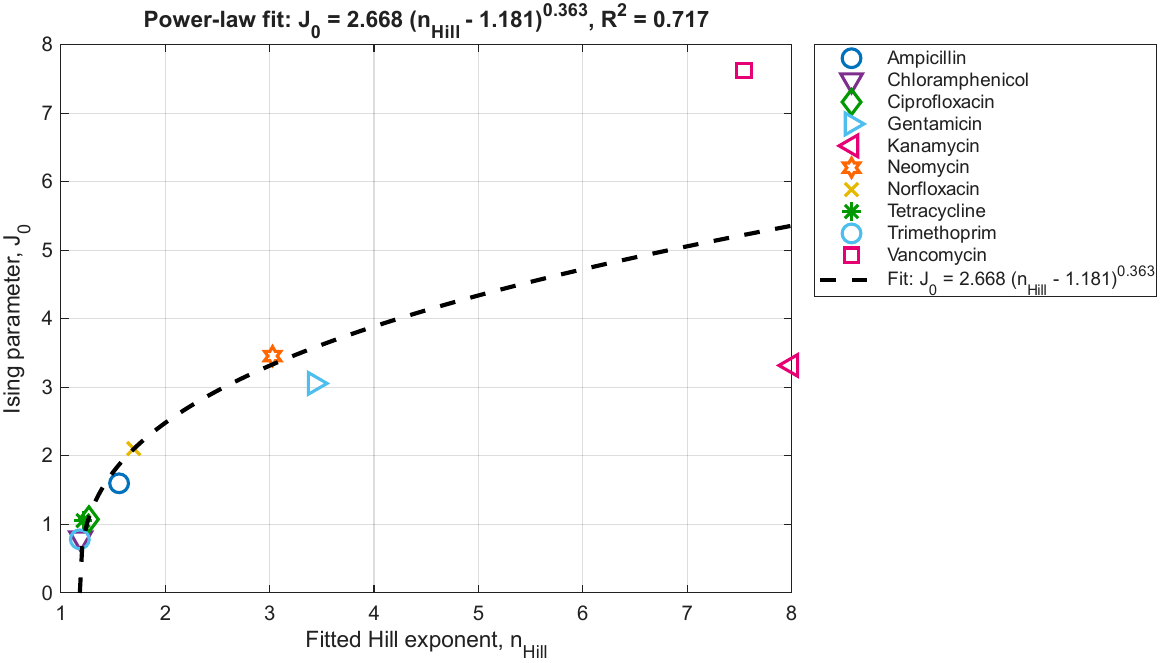}
        \caption{}
        \label{fig:panel_b}
    \end{subfigure}

    \vspace{0.1cm}

    % Panel (c)
    \begin{subfigure}{0.85\linewidth}
        \centering
        \includegraphics[width=0.7\linewidth]{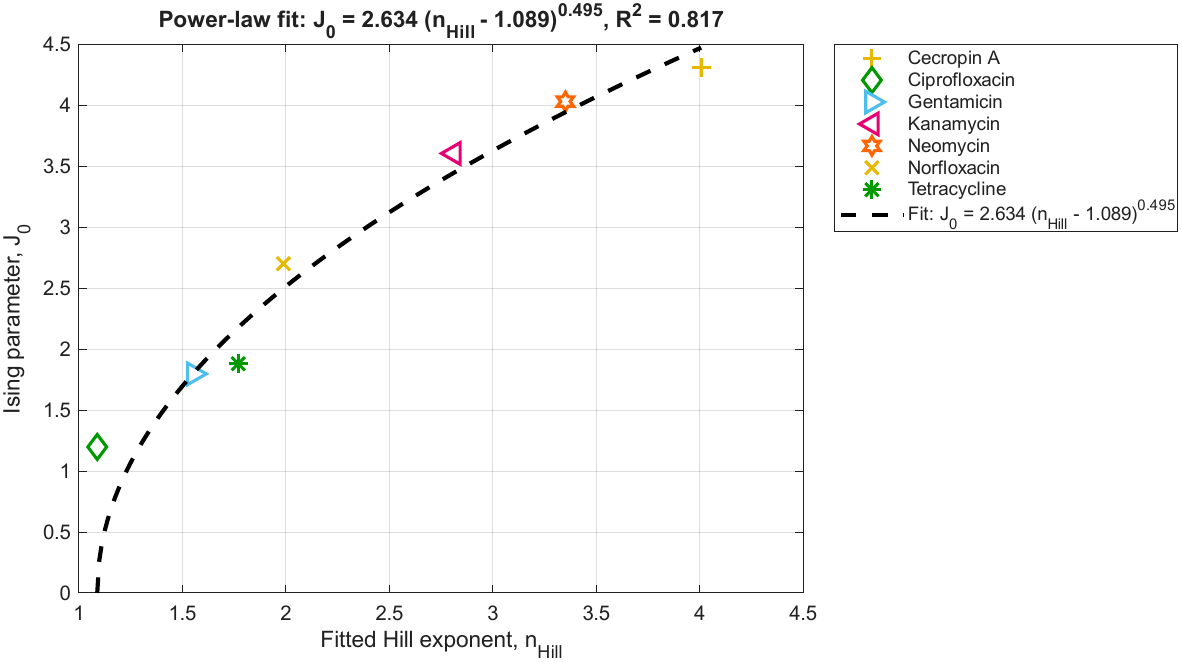}
        \caption{}
        \label{fig:panel_c}
    \end{subfigure}
\captionsetup{justification=raggedright,singlelinecheck=false}
    \caption{Comparison of fitted Hill exponents and Ising interaction strengths. (a) For \emph{E. coli}, the four black-filled symbols correspond to the antibiotics shown in Fig.1. The fits to the other antibiotics are shown in Figs.S4,S5. The dashed line is a polynomial best fit to serve as a guide of the general trend. (b) For \emph{Staph}, and (c) for \emph{Pseudomonas}.}

    \label{fig:ExponentJ}
\end{figure}

Comparison of fitted Hill exponents and Ising interaction strengths (Fig.\ref{fig:ExponentJ}).

\section{Fitting the Ising model to exact Hill function}

\begin{figure}[htbp]
    \centering

    % Panel (a)
    \begin{subfigure}{\linewidth}
        \centering
        \includegraphics[width=0.7\linewidth]{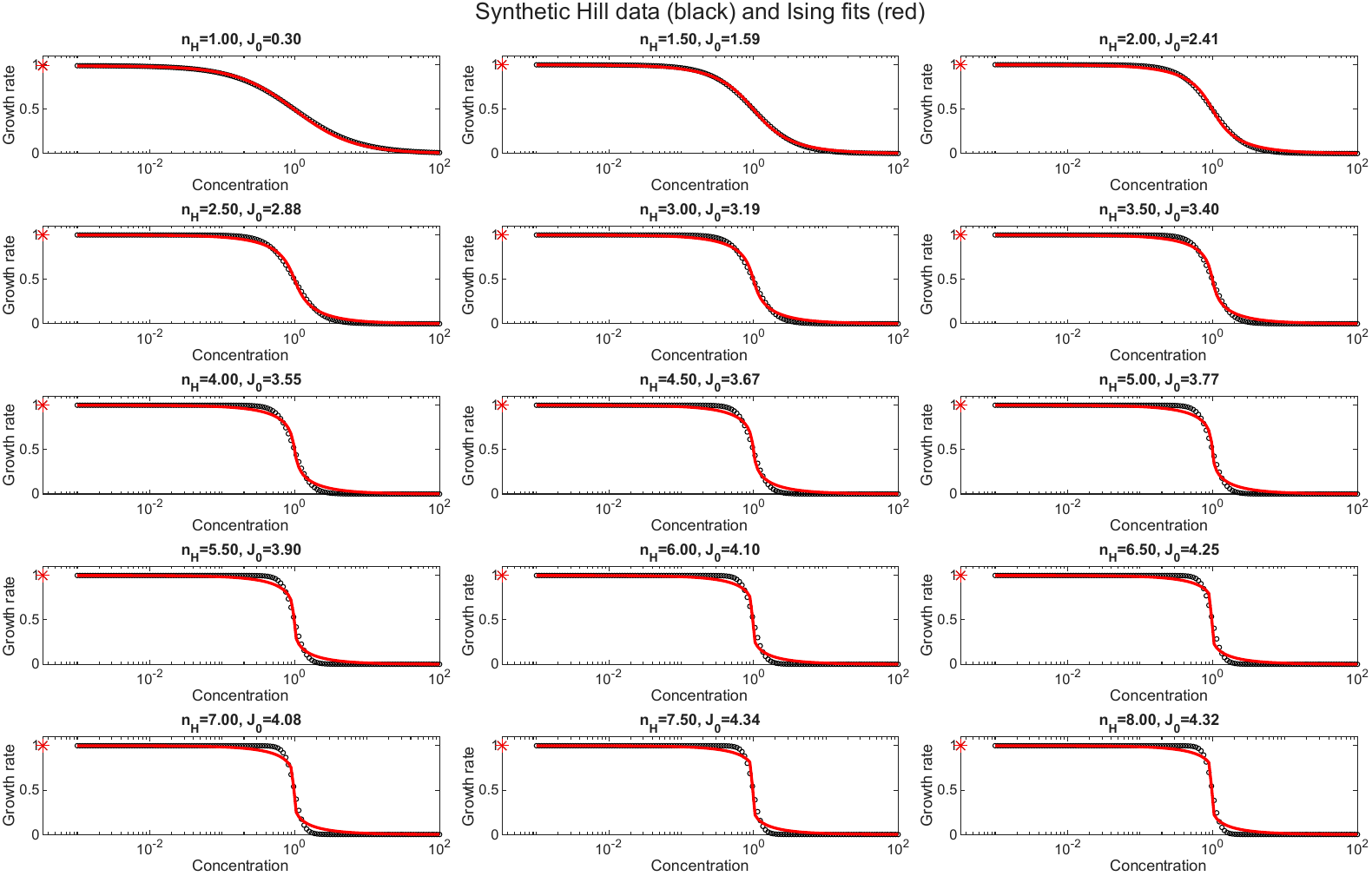}
        \caption{}
        \label{fig:panel_a}
    \end{subfigure}

    \vspace{0.1cm}

    % Panel (b)
    \begin{subfigure}{\linewidth}
        \centering
        \includegraphics[width=0.5\linewidth]{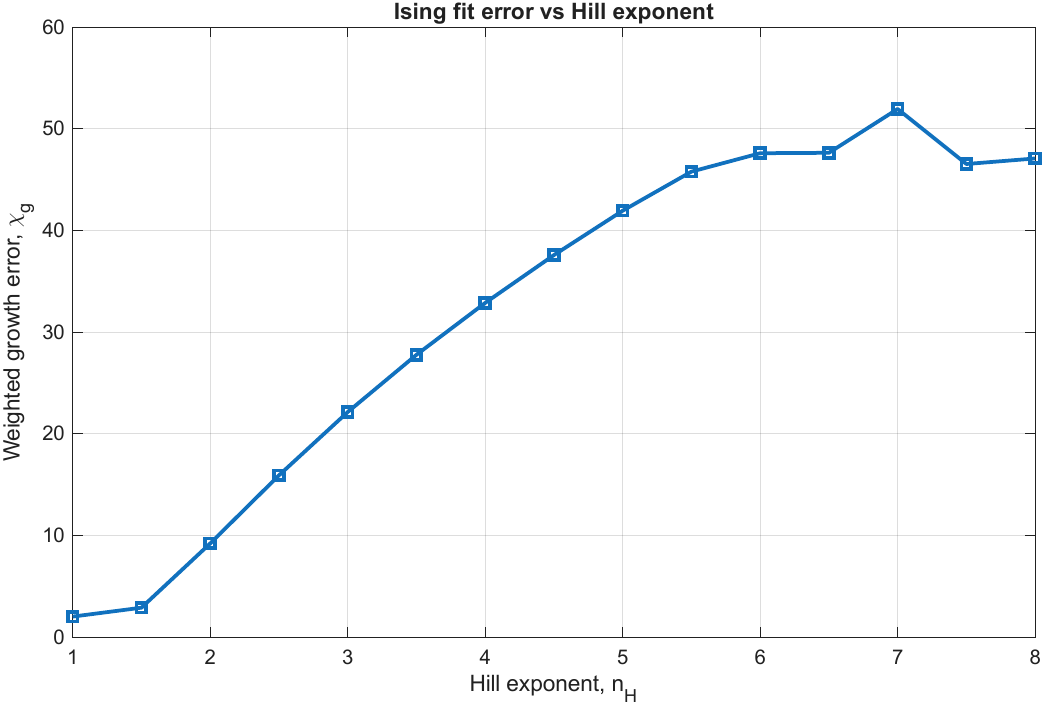}
        \caption{}
        \label{fig:panel_b}
    \end{subfigure}

    % Panel (c)
    \begin{subfigure}{\linewidth}
        \centering
        \includegraphics[width=0.7\linewidth]{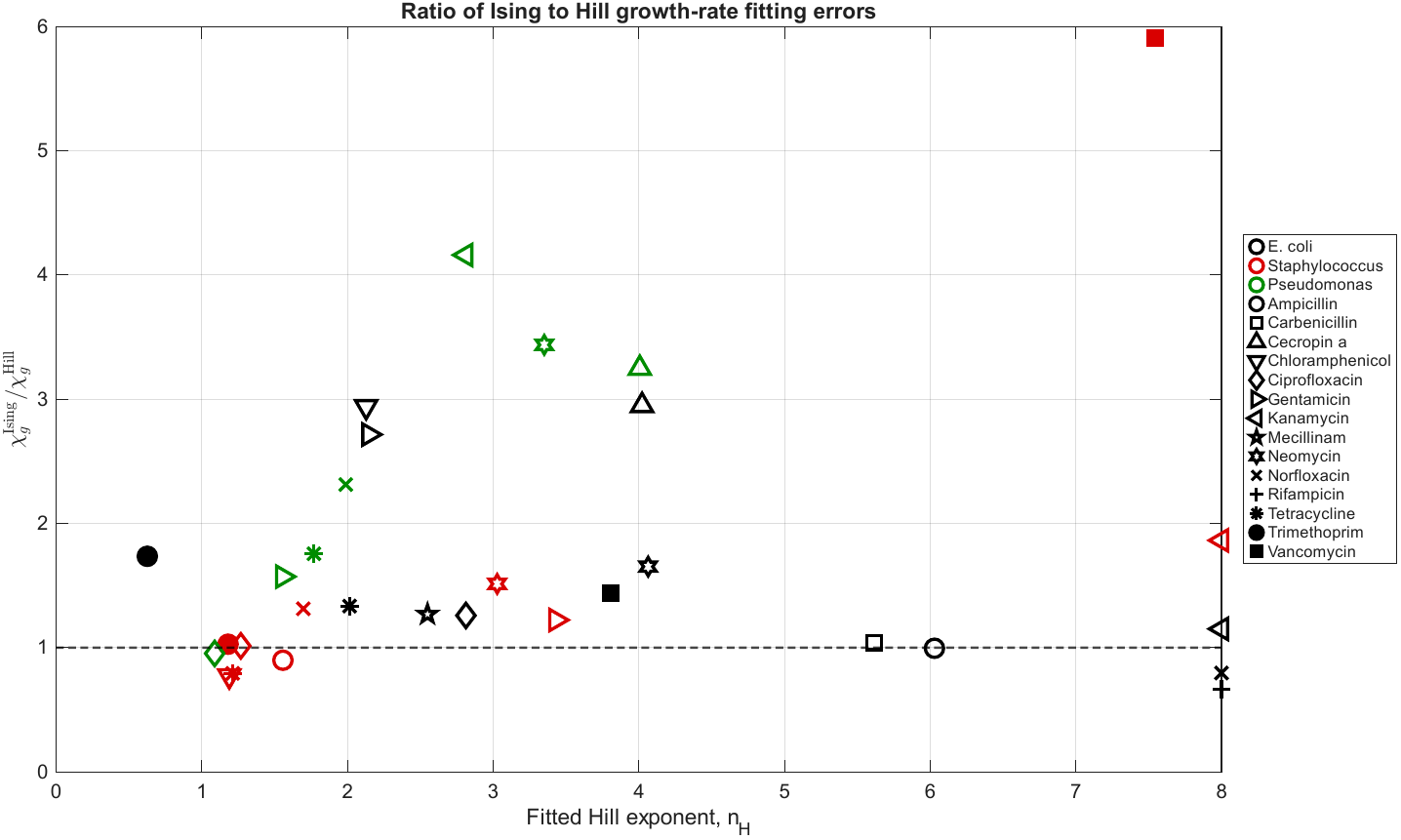}
        \caption{}
        \label{fig:panel_b}
    \end{subfigure}

%\captionsetup{justification=raggedright,singlelinecheck=false}
    \caption{ (a) Comparison of Hill functions (black circles) to fitting by an Ising model (red line), for different values of the Hill exponent. (b) The error of fitting the Ising model to the Hills function, as function of the Hill exponent. (c) Ratio of the Ising and Hill fitting errors to the growth-rate data as function of the Hill exponent (same data as in Fig.3).}

    \label{fig:synthetic}
\end{figure}

To get a better understanding of the relation between the Hill function exponent and the corresponding Ising model interaction, we produced data sets of exact Hill functions of different Hill exponent, densely and equally spaced along the logarithmic concentration axis. We then used the same procedure described in section S2 to fit it to the Ising model. The results are shown in Fig.\ref{fig:synthetic}, and the fitted values of $J_0$ are plotted as the blue dashed line in Fig.2. 

We see clearly that as the Ising model becomes sharper to match the Hill function for larger values of the exponent $n_H$, the two functions become more and more different as they approach the inflection point (Fig.\ref{fig:synthetic}a). This discrepancy is manifested in a monotonously increasing error for the quality of the Ising model fit to the Hill function (Fig.\ref{fig:synthetic}b). By comparison, if we plot the ratio of the Ising and Hill fitting errors to the growth-rate data as function of the Hill exponent (Fig.\ref{fig:synthetic}c), there is no monotonous increase. This may suggest that the experimental data is indeed not exactly Hill-like.

\section{Bistability and choice of the solution branch for Ising model fits with $J_0>J_c$.}

In Fig.2 (and Fig.\ref{fig:ExponentJ}) we plot the fitted Ising model interaction parameter $J_0$, and we find few cases where this value is significantly larger than the critical value of $J_c=4$. When this occurs, there is a large range of antibiotic concentrations for which there are two branches of solution of Eq.(11). We now inspect this bistability in more detail.

\begin{figure}[htbp]
    \centering

    % Panel (a)
    \begin{subfigure}{\linewidth}
        \centering
        \includegraphics[width=0.7\linewidth]{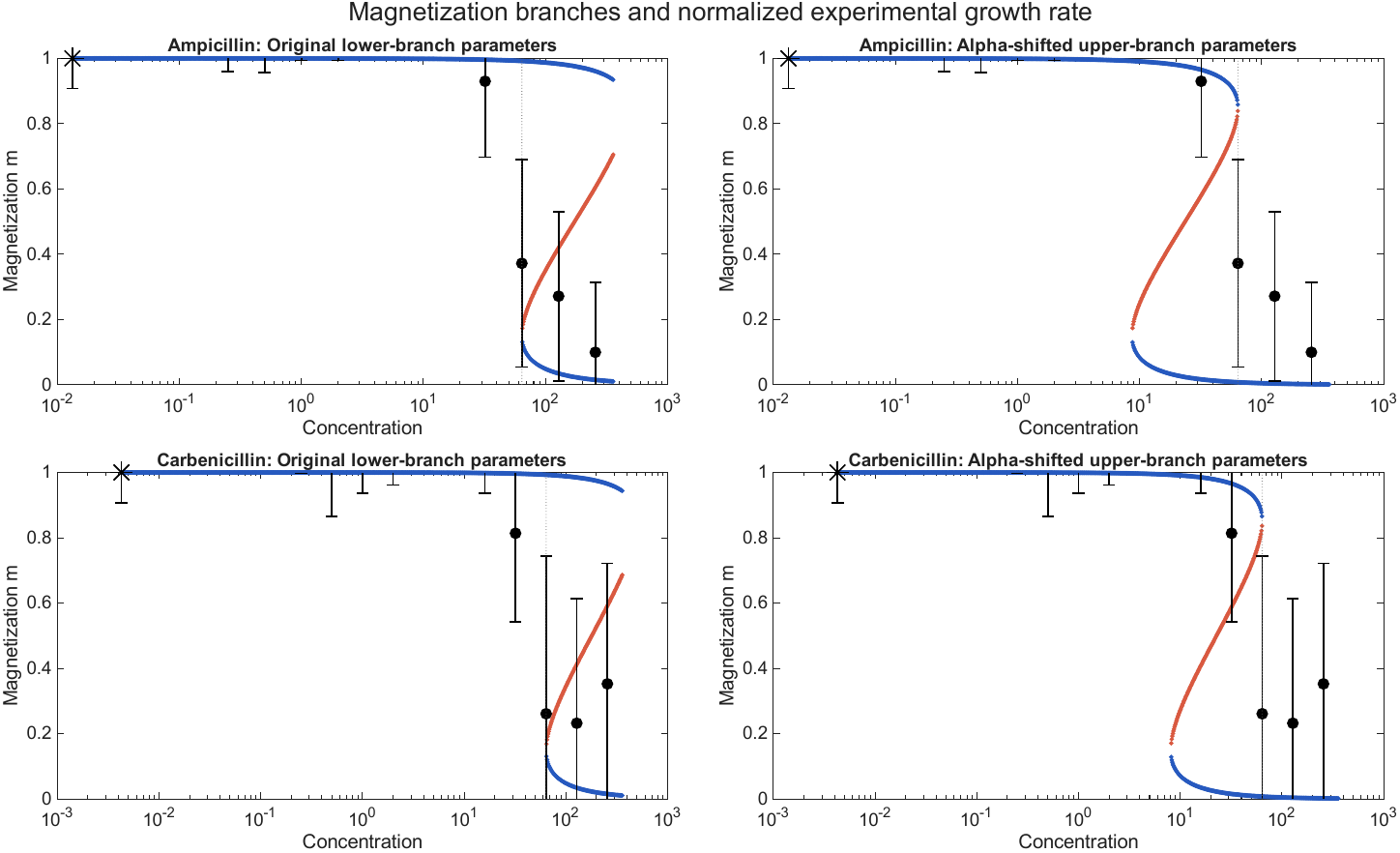}
        \caption{}
        \label{fig:panel_a}
    \end{subfigure}

    \vspace{0.1cm}

    % Panel (b)
    \begin{subfigure}{\linewidth}
        \centering
        \includegraphics[width=0.7\linewidth]{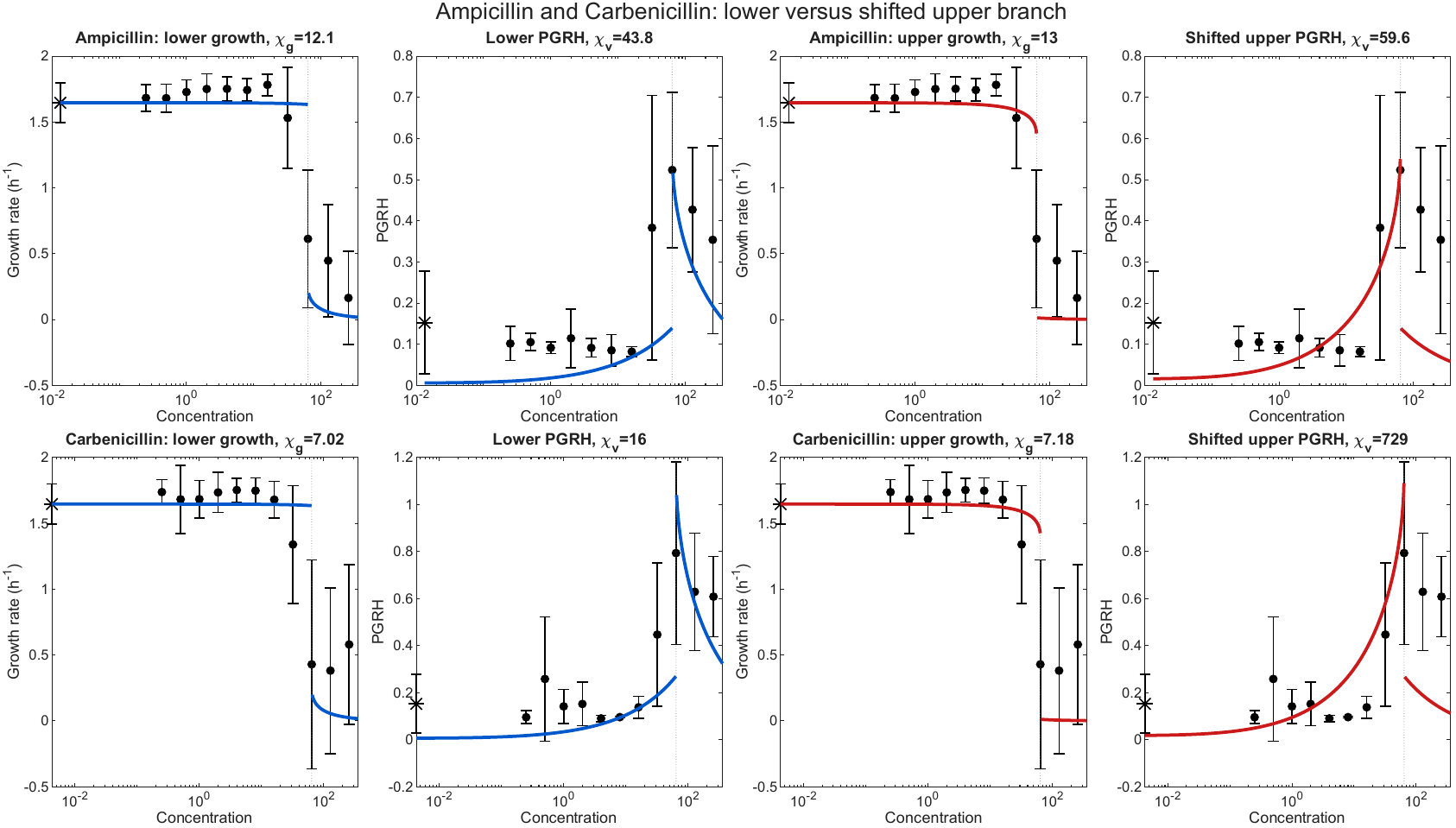}
        \caption{}
        \label{fig:panel_b}
    \end{subfigure}

%\captionsetup{justification=raggedright,singlelinecheck=false}
    \caption{ (a) Comparison of the growth-rate data of \emph{E. coli} as function of the concentration of Ampicillin and Carbenicillin (black circles and error bars). The lines show the fitted solution of the Ising model, including the bistable region. The left panels show the fit of the sharp drop (vertical dashed line) to the lower branch solution, and the right panels show the fits to the upper branch. Blue lines give the stable solutions and the red line denotes the unstable solution to Eq.(11). (b) Detailed comparison of the fitting of the two branches to the growth-rate data, and the corresponding comparison of the calculated variance (Eq.12) to the measured PGRH.}

    \label{fig:bistable}
\end{figure}

In Fig.\ref{fig:bistable}(a) we show this region by plotting the two branches of stable solutions, and the unstable solution, for the growth-rate of \emph{E. coli} as function of the concentration of Ampicillin and Carbenicillin. These antibiotics were fitted with very large values of $J_0=7.795,7.908$, respectively. 

In Fig.\ref{fig:bistable}(b) we show the detailed comparison of the fitting of the two branches to the growth-rate data, and the corresponding comparison of the calculated variance (Eq.12) to the measured PGRH. While the fitting to the growth-rate seems to be comparable using either one of the Ising solution branches (although they have distinctly different behavior on either side of the sharp drop), the lower branch (left panels, blue lines) seem to provide a much better fit for the PGRH. Specifically, the observed PGRH seems to have some of the asymmetric behavior predicted by the Ising model, with a low PGRH on the low concentration side of the drop, and much higher on the high concentration side (while its exactly opposite for the upper branch solution).

\section{Comparison of the errors of the fits of the Hills and Ising functions to the experimental data of the growth-rate and the PGRH.}

Comparison of the fitting errors for the Hill and Ising models, per each bacteria species, is given in Fig.\ref{fig:chi} (same data that was compiled together in Fig.3).

\begin{figure}[htbp]
    \centering

    % Panel (a)
    \begin{subfigure}{\linewidth}
        \centering
        \includegraphics[width=0.7\linewidth]{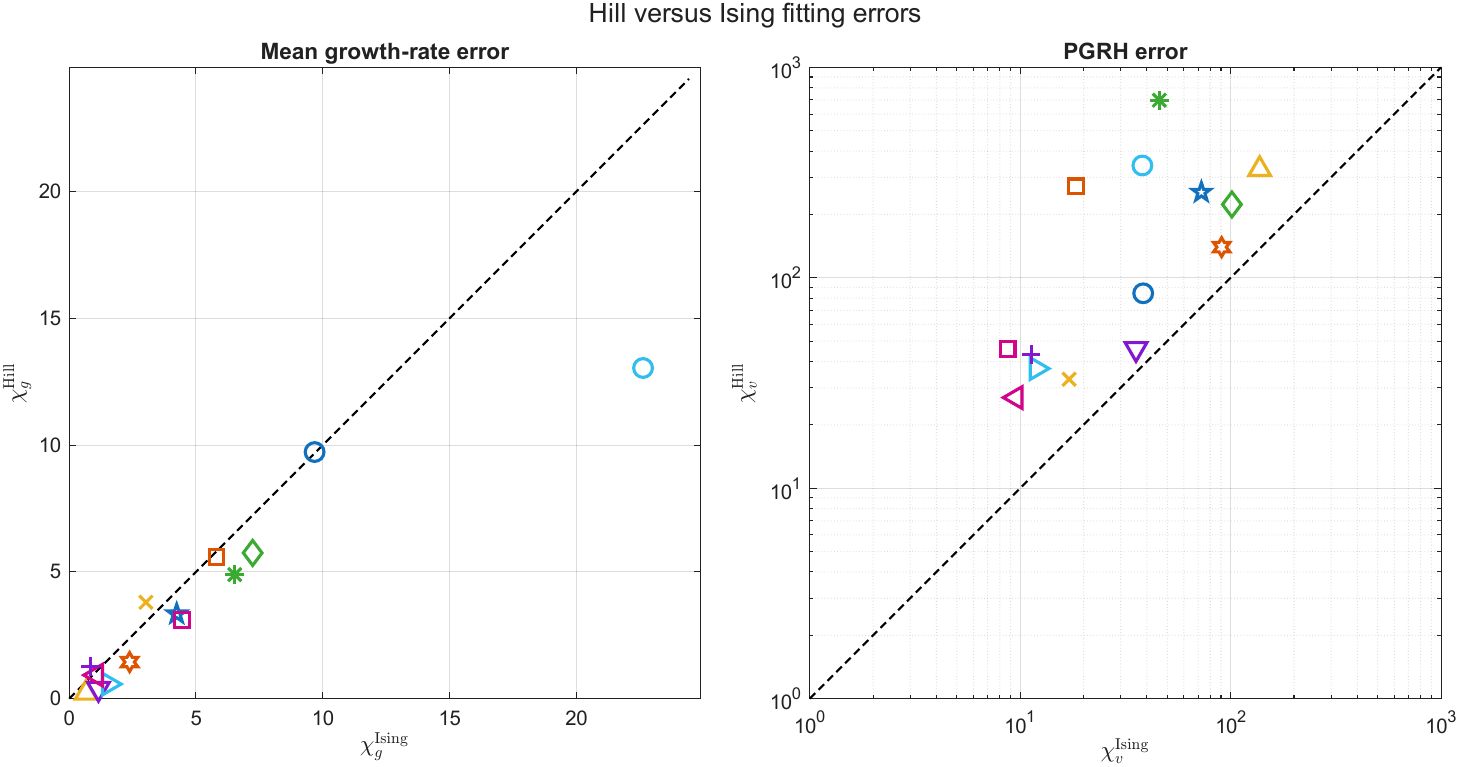}
        \caption{}
        \label{fig:panel_a}
    \end{subfigure}

    \vspace{0.1cm}

    % Panel (b)
    \begin{subfigure}{\linewidth}
        \centering
        \includegraphics[width=0.7\linewidth]{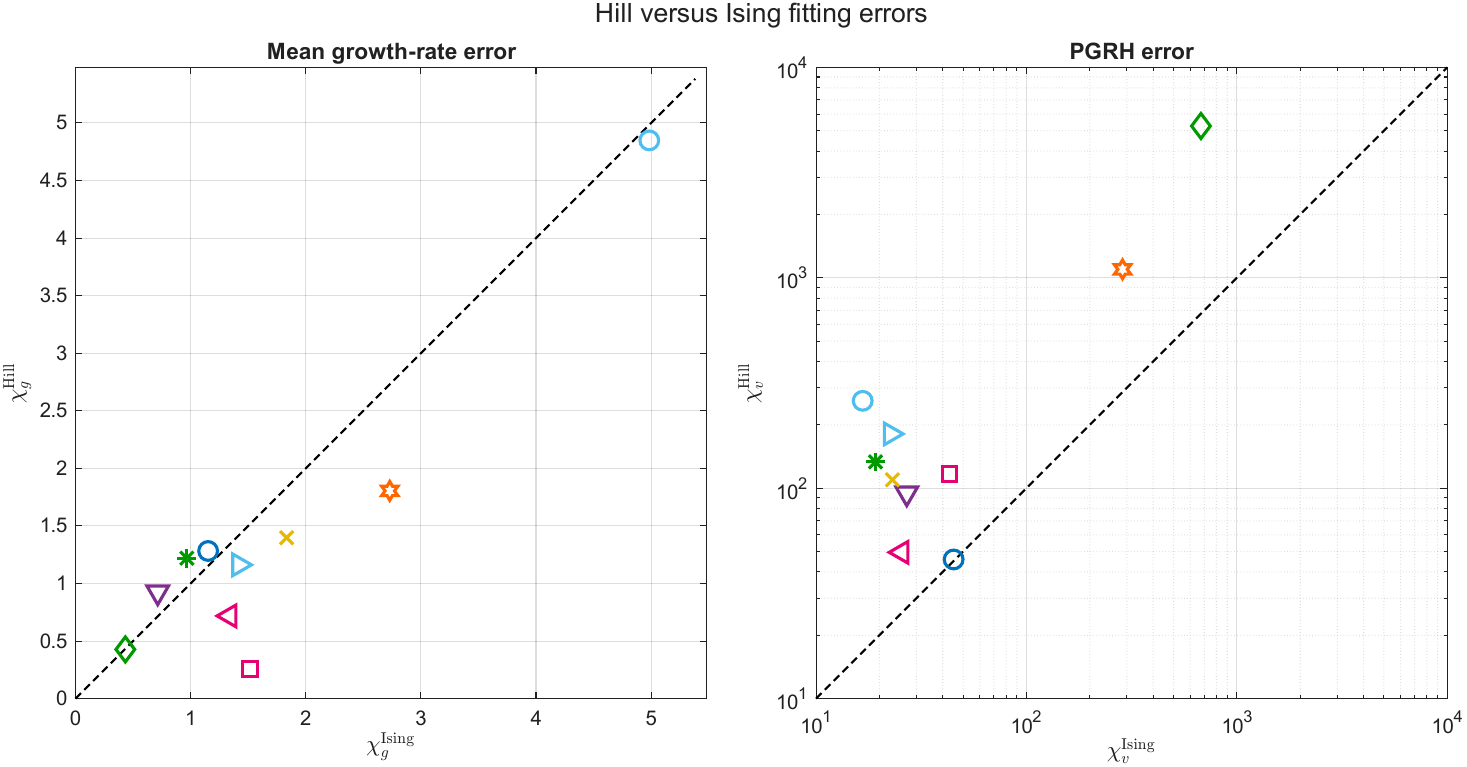}
        \caption{}
        \label{fig:panel_b}
    \end{subfigure}

    \vspace{0.1cm}

    % Panel (c)
    \begin{subfigure}{\linewidth}
        \centering
        \includegraphics[width=0.7\linewidth]{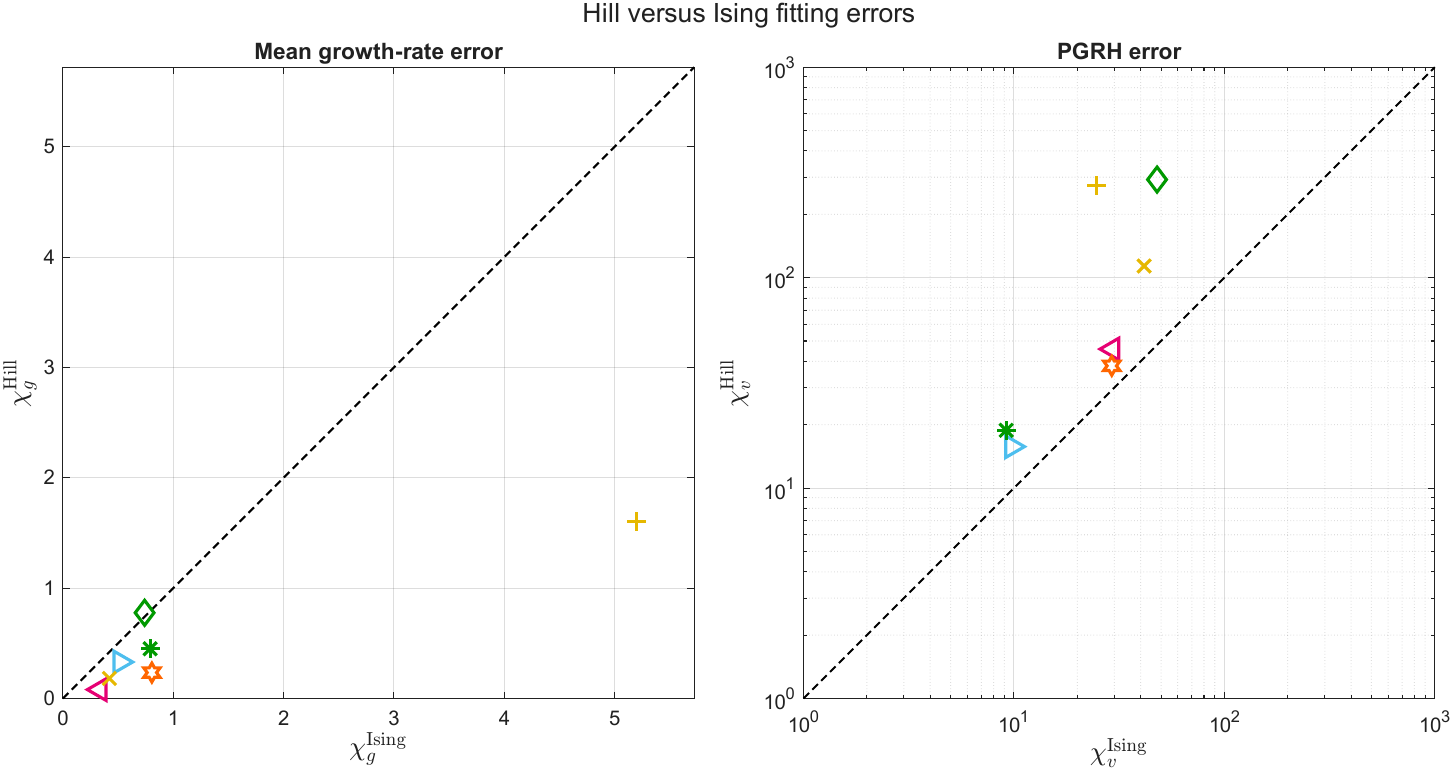}
        \caption{}
        \label{fig:panel_c}
    \end{subfigure}
%\captionsetup{justification=raggedright,singlelinecheck=false}
    \caption{ (a) Comparison of the fitting errors for the Hill and Ising models for \emph{E. coli}. The dashed line denotes the diagonal where the two errors are equal. Colors and symbols correspond to the different antibiotics as in Fig.\ref{fig:ExponentJ}. (b) For \emph{Staph}, and (c) for \emph{Pseudomonas}.}

    \label{fig:chi}
\end{figure}

\section{Concentration fluctuations that are proportional to the concentration, compared to the experimental data of the PGRH.}

We have so far considered that the antibiotic concentration fluctuations grow as the square-root of the concentration (Eq.2). This is motivated by Poisson statistics, which is most relevant at low concentrations. We therefore consider also the possibility that the concentration fluctuations may depend linearly on the concentration, and Eq.2 is modified to be
\begin{equation} \label{hillnoiseLin}
    \delta g(c)=A|g(c)'|c
\end{equation}
Where $A$ is an arbitrary multiplication factor.

\begin{figure}[htbp]
    \centering

    % Panel (a)
    \begin{subfigure}{\linewidth}
        \centering
        \includegraphics[width=0.7\linewidth]{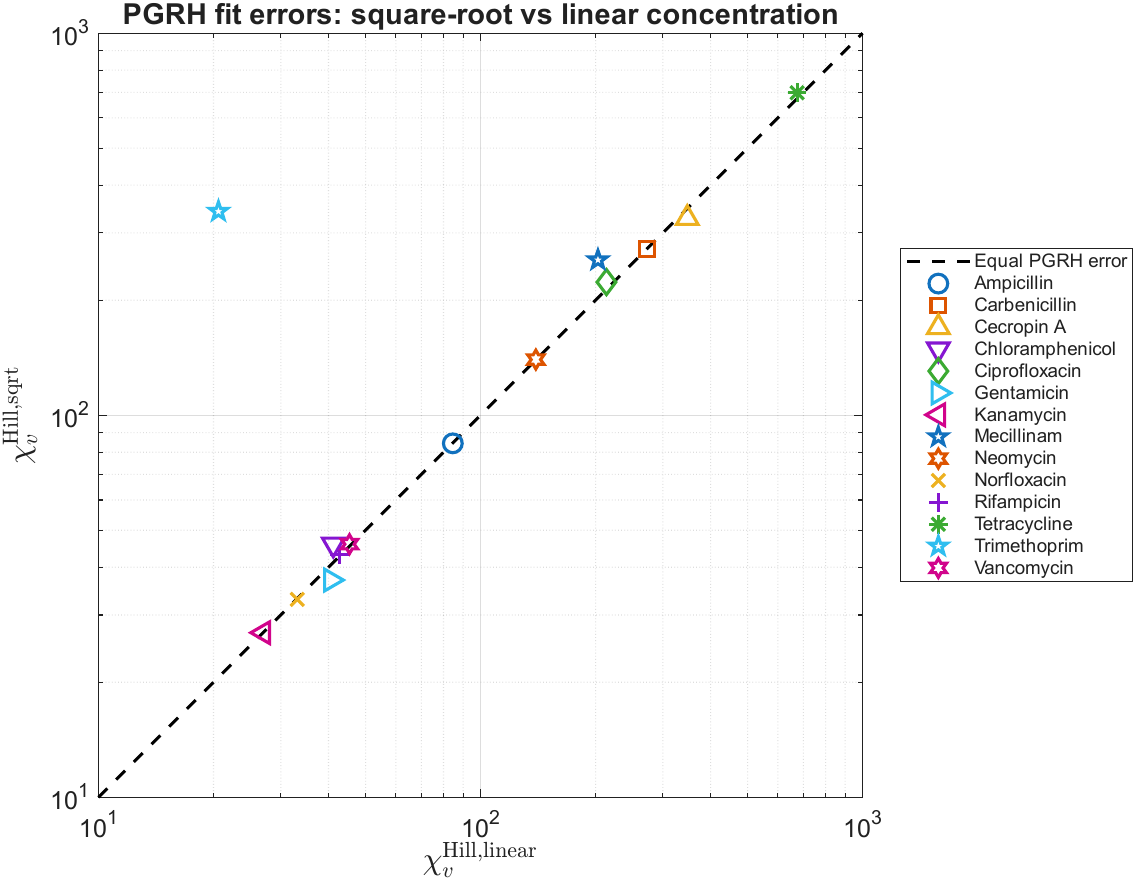}
        \caption{}
        \label{fig:panel_a}
    \end{subfigure}

    \vspace{0.1cm}

    % Panel (b)
    \begin{subfigure}{\linewidth}
        \centering
        \includegraphics[width=0.7\linewidth]%{Trimethoprim_HillFit_Linear.png}
        {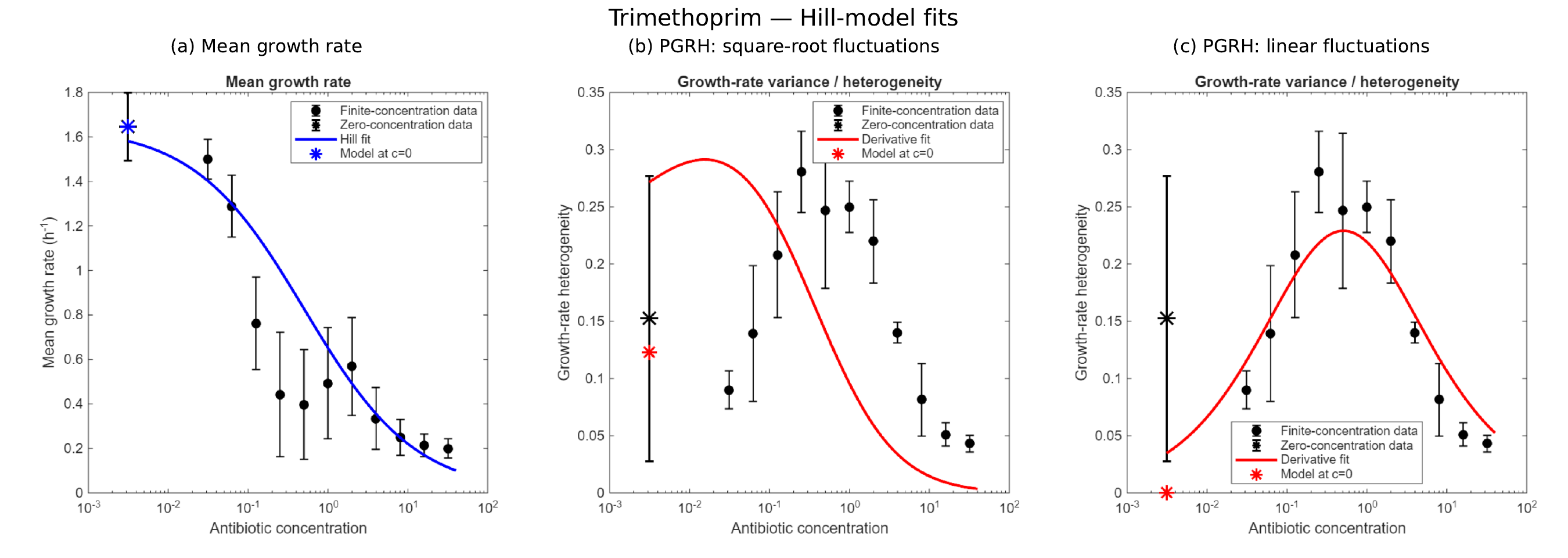}
        \caption{}
        \label{fig:panel_b}
    \end{subfigure}

    \vspace{0.1cm}

    % Panel (c)
\captionsetup{justification=raggedright,singlelinecheck=false}
    \caption{ (a) Comparison of the fitting errors for the PGRH, for \emph{E. coli}, between square-root and linear concentration fluctuations. The dashed line denotes the diagonal where the two errors are equal. (b) The Hill fit to the growth-rate (left) and to the PGRH (right panels), for square-root (Eq.2) and linear (Eq.\ref{hillnoiseLin}) concentration fluctuations (compare to Fig.\ref{fig:HillIsingSupplementary2}).}

    \label{fig:sqrtLinChi}
\end{figure}

In Fig.\ref{fig:sqrtLinChi} we plot the errors in the fits to the PGRH data (for \emph{E. coli}), using either the square-root or the linear relation. Except for one antibiotic (Trimethoprin, compare to Fig.\ref{fig:HillIsingSupplementary2}), there is no significant difference in the quality of the fits. 

\bibliography{references}

\end{document}